\documentclass[12pt]{iopart}

\pdfoutput=1
\usepackage{xspace}
\usepackage{color}
\usepackage{graphicx}
 
\renewcommand{\vec}{\mathbf}

\newcommand{\ybco}{YBa$_2$Cu$_3$O$_{7-\delta}$}

\newcommand{\BKFAop}{Ba$_{0.6}$K$_{0.4}$Fe$_2$As$_2$}
\newcommand{\BKFAops}{Ba$_{0.6}$K$_{0.4}$Fe$_2$As$_2$\xspace}

\begin{document}

\title[Fe-based superconductors: an ARPES perspective]{Fe-based superconductors: an angle-resolved photoemission spectroscopy perspective}

\author{P. Richard$^1$, T. Sato$^{2,3}$, K. Nakayama$^2$, T. Takahashi$^{2,4}$, H. Ding$^1$}

\address{1 Beijing National Laboratory for Condensed Matter Physics, and Institute of Physics, Chinese Academy of Sciences, Beijing 100190, China}
\address{2 Department of Physics, Tohoku University, Sendai 980-8578, Japan}
\address{3 TRiP, Japan Science and Technology Agency (JST), Kawaguchi 332-0012, Japan}
\address{4 WPI Research Center, Advanced Institute for Materials Research, Tohoku University, Sendai 980-8577, Japan}
\eads{\mailto{p.richard@iphy.ac.cn}}

\begin{abstract}
Angle-resolved photoemission spectroscopy allows direct visualization and experimental determination of the electronic structure of crystals in the momentum space, including the precise characterization of the Fermi surface and the superconducting order parameter. It is thus particularly suited for investigating multiband systems such as the Fe-based superconductors. In this review, we cover several aspects of these recently discovered materials that have been addressed by this technique, with a special emphasis on their superconducting gap and their Fermi surface topology. We provide sufficient experimental evidence to support the reliability and the consistency of the angle-resolved photoemission spectroscopy measurements over a wide range of material compositions. 
\end{abstract}

\maketitle

\section{Introduction}
Struggling for more than twenty years with a problem that it could not resolve despite tremendous efforts, the scientific community received in 2008 the announcement of the discovery of superconductivity at 26 K in a Fe-layered material \cite{Kamihara} as a powerful electro-shock. The monopoly of copper oxides (cuprates) on high-$T_c$ superconductivity was over. And yet, the future proved to be full of further surprises to those who started racing once more for the keys leading to higher $T_c$'s. 

Less than 4 years later, an impressive amount of data has been collected on these materials, accompanied by significant progress in our understanding of their exotic properties. Nevertheless, a consensus has not been reached on most aspects and it becomes very important to organize the pieces of the puzzle to get a more effective overall view. Up to now, there has been several review papers dealing with experiments on Fe-based superconductivity \cite{JohnstonAdv_Phys2010,GasparovJETP2010,PaglioneNPhys2010,StewartReview2011,LumsdenJPhys2010,MizuguchiJPSJ2010}. In the present paper, we do not intend to repeat the content of previous reviews, but rather to provide a different perspective based on a particular experimental technique, namely angle-resolved photoemission spectroscopy (ARPES). ARPES is a powerful tool to access \emph{directly} the momentum-resolved electronic structure of crystals, and thus it occupies a unique position when trying to understand their electronic behaviour. In this review, we will cover some important aspects of the Fe-based superconductors that have been addressed by ARPES, with a special emphasis on the superconducting (SC) gap.    

The review is organized as follows. We first provide basic knowledge to understand simply what ARPES is and what are its advantages and limitations. In Section \ref{section_structure}, we describe the electronic structure of the Fe-based superconductors, from core levels to the electronic states lying close to the Fermi level ($E_F$). The SC gap and its relationship to the electronic band structure will be discussed in Section \ref{section_gap}. We describe two models commonly used to explain superconductivity in these materials: the quasi-nesting model, which dominated the scene until only recently, and the local antiferromagnetic exchange pairing model, which now after the discovery of high-$T_c$ superconductivity in the 122-chalcogenides many consider a better candidate to unify the superconducting pairing mechanism in the Fe-based superconductors. Finally, before concluding this review, we will discuss the ARPES results in a broader context, where we compare ARPES with transport measurements.  

\section{Angle-resolved photoemission spectroscopy}
\label{section_ARPES}

\subsection{Basic principles}

\begin{figure}[htbp]
\begin{center}
\includegraphics[width=8cm]{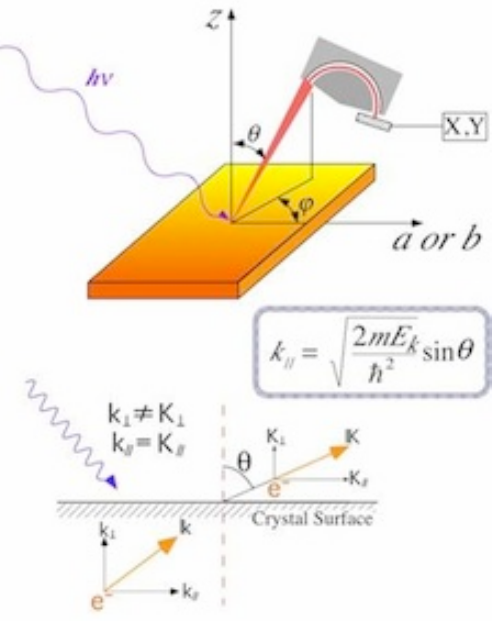}
\caption{\label{ARPES_process} (Colour online) Schematic diagram explaining ARPES measurements. Under a flux of photons of energy h$\nu$, electrons are ejected from the surface of a sample (orange slab) at an angle ($\theta$,$\varphi$) from the normal direction and detected with a high-resolution analyzer (in gray). The process implies the conservation of energy and in-plane momentum ($\vec{k}_{||}$).} 
\end{center}
\end{figure}

The idea of photoemission spectroscopy, which is simply based on the conservation of energy, is to extract information on the electronic structure of crystals by measuring the kinetic energy of electrons ejected from the surface of a sample due to the interaction with a photon flux of known energy h$\nu$ and vector potential $\vec{A}$. ARPES is a sophisticated method among a variety of photoemission spectroscopies. In addition to the conservation of energy, ARPES takes advantage of the conservation of the in-plane momentum. Hence, the direction of emission of the photo-excited electrons is recorded as well as their energy. This procedure, illustrated in Figure \ref{ARPES_process}, allows the determination of the electronic band dispersions. Although ARPES measurements require much more time than angle-integrated photoemission spectroscopy (PES), which provides information on the electronic density of states (DOS) only, they give access to a more comprehensive determination of the electric band structure. More specifically, the ARPES signal $I(\vec{k},E,\vec{A},h\nu)$ is proportional to the one-particle spectral weight $A$($\vec{k}$, $E$), which is the probability to have an electron in the sample with momentum $\vec{k}$ and energy $E$, times the Fermi-Dirac distribution $f$($E$,$T$): 

\begin{equation}
I(\vec{k},E,\vec{A},h\nu) = M(\vec{k},E,\vec{A},h\nu)A(\vec{k},E)f(E,T)
\end{equation}

\noindent where $M$ represents the photoemission matrix element determined by the photoemission process itself and carries no direct information on the band dispersion. However, $M$ contains precious information on the nature of the electronic states probed.

The ARPES experimental data can be represented by colour images, where the colour scale gives the photoemission intensity, which is generally displayed as a function of energy and momentum. The energy is usually referenced to the leading edge of a metallic film spectrum, typically of gold, in electric contact with the sample. Although colour images give a more intuitive and natural representation of the electronic dispersion, they fail to reveal the precise spectral lineshape. For this reason, it is also convenient to represent the ARPES data with a series of curves of photoemission intensity as a function of energy at a fixed momentum $k$, or energy distribution curves (EDCs). Alternatively, one can display a series of curves of photoemission intensity as a function of momentum at a fixed energy $E$, or momentum distribution curves (MDCs).

\subsection{Main advantages and limitations}
\label{advantages_limitations}

As with any experimental techniques, ARPES has advantages and limitations. Its main advantage is quite obvious: it allows a \emph{direct} determination of the electronic structure of materials, including the Fermi surface and the band dispersion in the momentum space. This is crucially important when studying multi-band systems like the Fe-based superconductors. The experimental information is irreplaceable since it is not obtained from a fitting procedure of several parameters on various models, which may turn out to be inadequate. Moreover, ARPES is suitable for relatively ``dirty" systems such as the copper oxides or the Fe-based superconductors, where doping is introduced through partial chemical substitution or the incorporation of interstitial dopants and vacancies. Such doping process induces disorder, making determination of Fermi surfaces from quantum oscillation measurements difficult or even impossible under low magnetic field. In contrast to transport techniques, ARPES is not limited to the Fermi surface, and it is sensitive to states away from the Fermi level, which is very useful when the physical properties are partly determined by high-energy states. Moreover, ARPES can be used to determine the orbital characters of the electronic states by taking advantage of selection rules involved in the photoemission process.   

The main disadvantage of ARPES, which can be also regarded as an advantage when dealing with surface phenomena as in the topological insulators, is its surface sensitivity. To avoid surface contamination, single-crystals of complex compounds are usually cleaved \emph{in situ} and measured in a vacuum better than 10$^{-9}$ Torr, the finite lifetime of the samples increasing as vacuum improves. Hence, the lifetime of the samples can vary from a few hours to a few days. To maximize the sample lifetime, vacuum in the 10$^{-11}$ Torr range are routinely achieved using modern ultra-high vacuum systems such as in the Institute of Physics of the Chinese Academy of Sciences or in Tohoku University. Strictly speaking, ARPES measures the surface of materials, not the bulk. Nevertheless, the surface is \emph{always} to some extend related to the bulk of materials. In other words, \emph{surface} = \emph{bulk} + $\delta$. The reliability of the ARPES measurements when interpreting the physical properties of the bulk is thus determined by the size of $\delta$, which is compound-dependent. There are several ways to check if $\delta$ is qualitatively small or large. $\delta$ can be considered small when:

\begin{enumerate}
\item Low energy electron diffraction (LEED) pictures do not show obvious surface reconstruction;
\item The surface carrier doping, as determined from the Luttinger theorem, is consistent with that of the bulk;
\item The Fermi surface evolves smoothly with doping.
\item The electronic structure (band dispersion, gap size, etc...) varies with $k_z$, in sharp contrast to pure surface states;
\item The SC gap observed by ARPES closes at the bulk $T_c$;
\item The core levels of the relevant elements are not doubled;
\item The band dispersions are similar, albeit for some renormalization, to LDA predictions;
\item No unexpected band folding is observed by ARPES;
\item Gaps measured by ARPES are consistent with gaps measured from bulk-sensitive probes. It is to note that ARPES bulk-sensitivity is highly enhanced by the use of very low photon energies ($h\nu<9$ eV) \cite{KissPRL2005,SoumaRSI2007} or high photon energies ($h\nu$ > 500 eV) \cite{KamakuraEPL2004}.
\end{enumerate}

Fortunately, several Fe-based superconductors fulfill these requirements. Although the influence of the surface cannot be predicted systematically, the crystal structure symmetry of layered systems guides us to determine which compound is susceptible to present a surface problem. When the cleaving plane is unique and occurs between two symmetric layers, the two cleaved pieces are perfectly symmetric. The surface is thus usually non-polar and surface effects are expected to be minimal. For example, that case corresponds to the 11 and 111 structural phases of Fe-based superconductors. On the other hand, when the structure presents more than one equivalent cleaving planes leading to more than one possible surface exposed, the surface may be highly polar. The surface doping in this case is not representative of the bulk. YBa$_2$Cu$_3$O$_{7-\delta}$ is one of the most famous structures belonging to this category, in which the 1111-pnictide structure falls as well. Nevertheless, ARPES can be used in this case to get a reasonable idea of the band structure, exact doping and $k_z$ dispersion put aside. Another important situation is the one when cleaving occurs exactly at a symmetry plane occupied by one layer of atoms. To account for electrostatic stability, only half of the atoms remain on the cleaved surface. The surface may reconstruct or not. The experiment alone and the use of the criteria described above can determine whether the experimental results are representative of the bulk for the electronic states of interest. A special attention need to be given to the 122 family of Fe-pnictides, which belongs to this category. Surface effects have been reported in SrFe$_2$As$_2$ \cite{Hsieh}. Interestingly, such effect is not observed in hole-doped Ba$_{1-x}$K$_x$Fe$_2$As$_2$, which satisfies the criteria enumerated above, in agreement with a LEED and STM study reporting the absence of surface reconstruction in BaFe$_2$As$_2$ \cite{Nascimento_PRL2009}. Furthermore, as shown below, the 122-pnictides share similar electronic structures and SC gap functions with other Fe-based superconductors with ideal cleaving planes, suggesting that even supposing that a small distortion occurs in the (Ba, K) layer at the cleaved surface, the effect on the Fe electronic states are negligible.

\subsection{Brillouin zone notation}  

Unfortunately, the notation used in ARPES studies of Fe-based superconductors to describe high-symmetry points in the momentum space is not homogeneous, and two main representations are usually found in the literature. In the first one, only the Fe atoms are considered. The Fe atoms form a square lattice of parameter $a$, which corresponds to a square Brillouin zone (BZ) of size $2\pi/a$. In this \emph{1 Fe/unit cell} or \emph{unreconstructed} description, the zone center $\Gamma$ and the M point correspond to $(0,0)$ and $(\pi/a,0)$, respectively, while the X point is defined as $(\pi/2,\pi/2)$. The $\Gamma$-M direction is associated to momentum along the Fe-Fe bounding. Alternatively, one can prefer to consider the ``real" BZ and take into account the effect of the alternative distribution of chalcogen and pnictogen atoms up and down the Fe plane. The unit cell parameter becomes $a'=a\sqrt{2}$. In this \emph{2 Fe/unit cell} or \emph{reconstructed} description, M = $(\pi/a',\pi/a')$ and X = $(\pi/a',0)$. To add to the confusion, some authors rather define M = $(\pi/a',0)$ and X = $(\pi/a',\pi/a')$. Unless specifically indicated, we use the 1 Fe/unit cell representation by default in the present review.

\section{Electronic structure}
\label{section_structure}

The electronic properties of a material are governed by its own electronic structure, which is determined by the composition and the arrangement of the atoms from which it is made. All Fe-based superconductors are characterized by the presence of Fe layers. It is thus quite natural that they share similar electronic structure. Since it has been studied in more details, we begin this chapter by presenting ARPES results on optimally-doped \BKFAops as a typical Fe-based superconductor. We will then review ARPES results obtained on several other Fe-based materials.

\subsection{Optimally-doped Ba$_{0.6}$K$_{0.4}$Fe$_2$As$_2$}  

\begin{figure}[htbp]
\begin{center}
\includegraphics[width=8cm]{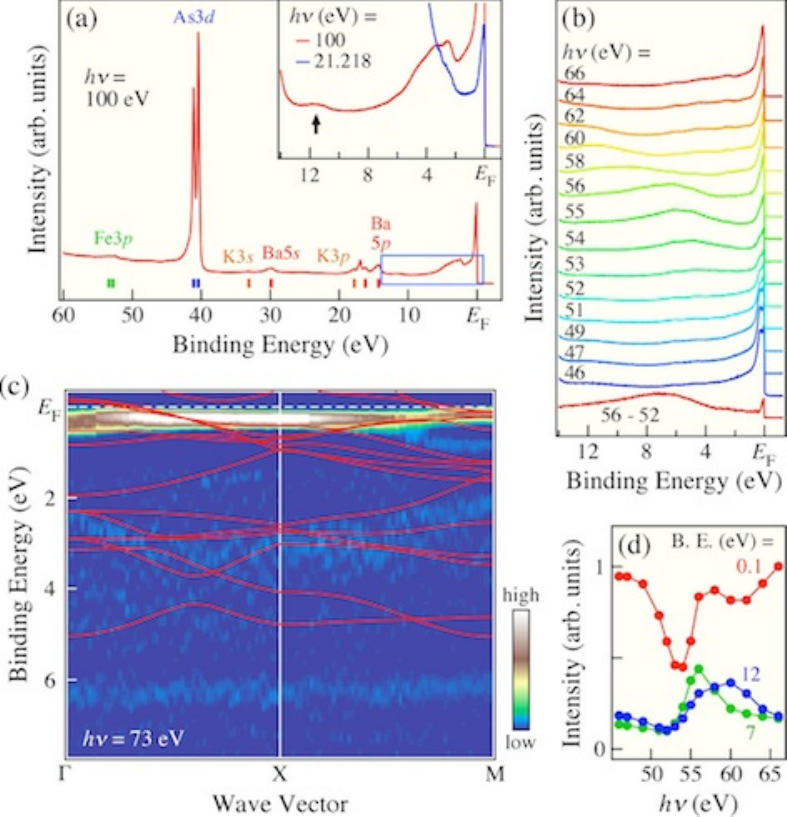}
\caption{\label{core_levels_OP} (Colour online)  (a) Wide range EDC near $\Gamma$ showing shallow core levels marked by vertical bars above the $x$-axis. The inset magnifies the valence band and a possible satellite peak at $\sim$ 12 eV, and highlights the difference between spectra taken at 100 and 21.2 eV. (b) Valence band near $\Gamma$ measured at different photon energies (46 - 66 eV). All EDCs are normalized by the photon flux. (c) Intensity plot of second derivatives of spectra along $\Gamma$-X and X-M. LDA bands (red lines) are plotted for comparison. (d) Photon energy dependence of the EDC intensity shown in (b) obtained at binding energies 0.1, 7, and 12 eV. From Ding \emph{et al.}, J. Phys.: Condens. Matter., \textbf{23}, 135501 (2011) \cite{Ding_J_Phys_Condens_Matter2011}, copyright \copyright\xspace (2011) by IOP Publishing.} 
\end{center}
\end{figure}

The core level spectrum displayed in Figure \ref{core_levels_OP}(a) is consistent with the chemical composition of \BKFAop. From high to low binding energies, we can distinguish unambiguously core levels associated with Fe$3p$ (52.4, 53.0 eV), As$3d$ (40.4, 41.1 eV), K$3s$ (33.0 eV), Ba$5s$ (29.7 eV), K$3p$ (17.8 eV) and Ba$5p$ (14.2, 16.2 eV) \cite{Ding_J_Phys_Condens_Matter2011}. We also observe a weak peak around 12 eV, whose origin remains unclear. Although it could come from As$4s$ states, we note that several divalent Fe compounds, such as FeO, exhibit a peak at similar energy attributed to a satellite state of the Fe$3d^5$ configuration \cite{Lad_PRB1989}. The strongest core level peaks are those corresponding to the As$3d_{5/2}$ and As$3d_{3/2}$ levels. Although As atoms are located just below the cleaved surface, it is worth mentioning that unlike GaAs, for which a set of surface components are observed 0.4 eV below the peak energies, the As $3d$ levels in \BKFAops do not show obvious additional component that could result from a surface reconstruction. It is thus reasonable to assume that the electronic states of Fe atoms, which are located below the first As layer, are not affected significantly by the surface termination. 

The nature of the electronic states within about 8 eV below $E_F$ can be determined by investigating their photon energy dependence. The inset of Figure \ref{core_levels_OP}(a) compares the spectra of \BKFAops at 21.2 eV and 100 eV. While the former photon energy is more sensitive to As$4s$ states, the latter enhances Fe$3d$ states. We thus conclude that the sharp peak within 1 eV below $E_F$ originates mainly from Fe$3d$ states and the states at higher biding energy mainly from As$4s$ orbitals, in agreement with LDA band calculations \cite{Ma_Front_phys_China}. This assessment is reinforced by the detailed photon energy evolution of the low energy states across the Fe$3p\rightarrow$Fe$3d$ resonance at 56 eV, which is illustrated in Figure \ref{core_levels_OP}(b). The difference of the spectra recorded at 56 eV (at the resonance) and at 52 eV (below the resonance) exhibits two features: a sharp peak near $E_F$ and a broad peak centered around 7 eV, which can be regarded as the coherent and incoherent parts of the Fe$3d$ states, respectively. While the photoemission intensity near $E_F$ shows an anti-resonance profile, the incoherent part at 7 eV has a Fano-like resonance profile (see Figure \ref{core_levels_OP}(d)), which is similar to the Fe$3d$ states in FeO \cite{Lad_PRB1989}. 

The Fe$3d$ and As$4s$ levels within 8 eV below $E_F$ are dispersive, as illustrated in Figure \ref{core_levels_OP}(c) by the intensity plot of second derivative, for the $\Gamma$-M and $\Gamma$-X high symmetry lines. When renormalized by a factor 2, LDA band calculations \cite{G_Xu_EPL2008} at this doping capture some overall features observed experimentally. This indicates that correlations in this material are not negligible. Interestingly, band renormalization seems more important at lower binding energy, as discussed below. 
  
\begin{figure}[htbp]
\begin{center}
\includegraphics[width=8cm]{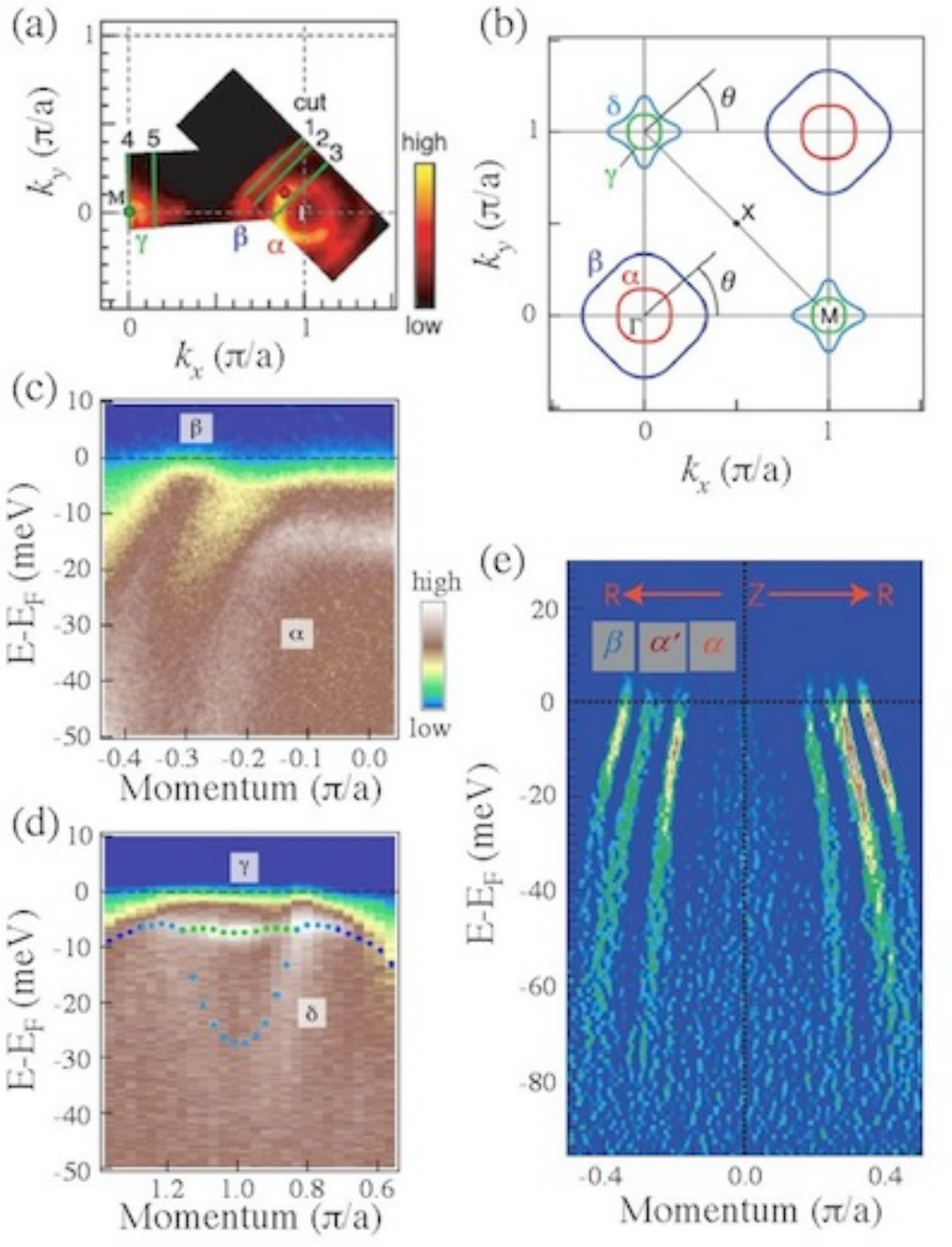}
\caption{\label{FS_OP} (Colour online) (a) FS contour determined by plotting the ARPES spectral intensity integrated within $\pm 10$ meV with respect to $E_F$. (b) Schematic view of the four FS sheets with a definition of the FS angle ($\theta$). (c) Intensity plot near $\Gamma$ measured at $T=15$ K. (d) Intensity plot near M measured at 15 K. Dots are EDC peak positions. (e) Second derivative plot of the dispersion along Z-R ($k_z\pi$) measured using 32 eV photons. Three hole-like bands ($\alpha$ (inner), $\alpha$' (middle) and $\beta$ (outer)) are observed. (a) is from Ding \emph{et al.}, Europhy. Lett., \textbf{83}, 47001 (2008) \cite{Ding_EPL}, copyright \copyright\xspace (2008) by the European Physical Society. (b) is from Nakayama \emph{et al.}, Europhy. Lett., \textbf{85}, 67002 (2009) \cite{Nakayama_EPL2009}, copyright \copyright\xspace (2009) by the European Physical Society. (c) and (d) are from Ding \emph{et al.}, J. Phys.: Condens. Matter., \textbf{23}, 135501 (2011) \cite{Ding_J_Phys_Condens_Matter2011}, copyright \copyright\xspace (2011) by IOP Publishing. (e) is from Xu \emph{et al.}, Nature Phys., \textbf{7}, 198 (2011) \cite{YM_Xu_NPhys2011}, copyright \copyright\xspace (2011) by Macmillan Publishers Ltd.} 
\end{center}
\end{figure}

We now turn our attention to the states closer to the Fermi energy, which mainly govern the electronic properties of materials. The Fermi surface of \BKFAops recorded with the He I$\alpha$ resonance line is displayed in Figure \ref{FS_OP}(a) \cite{Ding_EPL}, and a schematic version is given in Figure \ref{FS_OP}(b). It illustrates the multi-band nature of the Fe-based superconductors. ARPES intensity plots along cuts passing through or close to the $\Gamma$ point allow us to identify at least two holelike bands crossing $E_F$. Hereafter, we call $\alpha$ and $\beta$ the bands giving rise to the small and large $\Gamma$-centered FS pockets, respectively. These band are quite clear in the intensity plot given in Figure \ref{FS_OP}(c) (below $T_c$). Similarly, the ARPES intensity plots along cuts passing near the M point reveal two distinct electronlike FS pockets, $\gamma$ and $\delta$, as shown in Figure \ref{FS_OP}(d). Their origin can be understood from the hybridization of ellipses elongated towards the $\Gamma$-M direction and folded across the $\Gamma$-X plane due to the BZ reconstruction induced by the alternative positions of As atoms below and above the Fe layer. 

Although very high energy resolution can be achieved in ARPES measurements performed with a He discharging lamp, the results are confined to a single $k_z$ value. Using the tunability of synchrotron radiation, ARPES experiments can reveal a small warping in the band dispersion along $k_z$ \cite{YM_Xu_NPhys2011,Y_Zhang_PRL2010}. More importantly, an additional holelike band, which is almost degenerate with the $\alpha$ band at the zone center, is resolved as $k_z$ increases towards Z $=(0,0,\pi/c)$ (where $c=6.5$ \AA\xspace is the lattice parameter of the primitive unit cell and the distance between two Fe layers), as illustrated in Figure \ref{FS_OP}(e). The synchrotron data indicate that He I$\alpha$ radiation corresponds approximately to $k_z=0$, explaining why this third band was not observed in the earliest ARPES experiments. The results obtained at different values of $k_z$ have several significant consequences: (i) They are consistent with LDA band calculations, which all predict 3 $\Gamma$-centered holelike bands and 2 M-centered electronlike bands; (2) They also confirm that the surface of \BKFAops does not carry extra charge, unlike \ybco. Indeed, assuming a double degeneracy for the $\alpha$ band, one can show that the algebraic FS area satisfies the Luttinger theorem with the bulk doping \cite{Ding_EPL, Nakayama_EPL2009}; (iii) Finally, the $k_z$ warping of the electronic structure proves that the electronic states measured by ARPES \emph{cannot} be pure surface states and that ARPES captures at least the essence of the bulk electronic properties.    

It is interesting to note that the Fermi velocities determined from ARPES differ from the values predicted by LDA calculations by a factor greater than the renormalized factor of 2 obtained for the overall band structure, indicating that low-energy states are further renormalized \cite{Ding_J_Phys_Condens_Matter2011}. Interestingly, such observation has also been done for other strongly correlated electron systems like the cuprates \cite{Valla_Science1999} and the cobaltates \cite{HB_YangPRL2005}. For practical purposes, the band structure can be parametrized using a tight-binding-like model \cite{Ding_J_Phys_Condens_Matter2011}: 

\begin{equation}
E^{\alpha,\beta}(k_x,k_y)= E_0^{\alpha,\beta}+t_1^{\alpha,\beta}(\cos k_x+\cos k_y)+t_2^{\alpha,\beta}\cos k_x\cos k_y
\end{equation}

\noindent and

\begin{equation}
E^{\gamma,\delta}(k_x,k_y)= E_0^{\gamma,\delta}+t_1^{\gamma,\delta}(\cos k_x+\cos k_y)+t_2^{\gamma,\delta}\cos k_x/2\cos k_y/2
\end{equation}

\noindent Since ARPES can reveal the unoccupied part of the band structure for only a few $k_BT$'s, mild constrains must be used for the fit. The results obtained with this model allow an estimation of the effective masses and reveal a total renormalization factor of about 3-4 for the near-$E_F$ bands, in addition of providing essential parameters to calculate some thermodynamical parameters such as the Sommerfeld coefficient \cite{Ding_J_Phys_Condens_Matter2011}.  

\begin{center}
\begin{table}[htbp]
\caption{\label{tablefit} Tight-binding fit parameters. From ref. \cite{Ding_J_Phys_Condens_Matter2011}.}
\begin{indented}
\item[]\begin{tabular}{@{}ccccc}
\br
&$\alpha$&$\beta$&$\gamma$&$\delta$\\
\mr
$E_0$&-0.24&-0.025&0.7&0.7\\
$t_1$&0.16&0.013&0.38&0.38\\
$t_2$&-0.052&0.042&0.8&-0.8\\
\br
\end{tabular}
\end{indented}
\end{table}
\end{center}

\subsection{Magnetic parent compounds}  

The parent compounds of both the cuprates and the Fe-based superconductors have magnetically ordered states, with superconductivity emerging away from the magnetic ordered phase. The study of these materials is therefore of prior interest. Unlike the cuprates though, the parent compounds of the Fe-based superconductors are generally metallic, which is an important advantage for ARPES studies. Hence, ARPES investigations have been reported on the parent compounds of the 122-pnictide \cite{Hsieh,VilmercatiPRB2009, LiuPRL2009, RichardPRL2010, FinkPRB2009, Malaeb_JPSJ2009, M_YiPRB2009_2,LX_YangPRL2009,GD_LiuPRB2009,Brouet_PRB2009,Kondo_PRB2010,M_YiPNAS2011,Y_KimPRB2011}, 11-chalcogenide \cite{Y_XiaPRL2009,Y_Zhang_PRB2010} and 111-pnictide \cite{C_HePRL2010} systems.

\begin{figure}[htbp]
\begin{center}
\includegraphics[width=8cm]{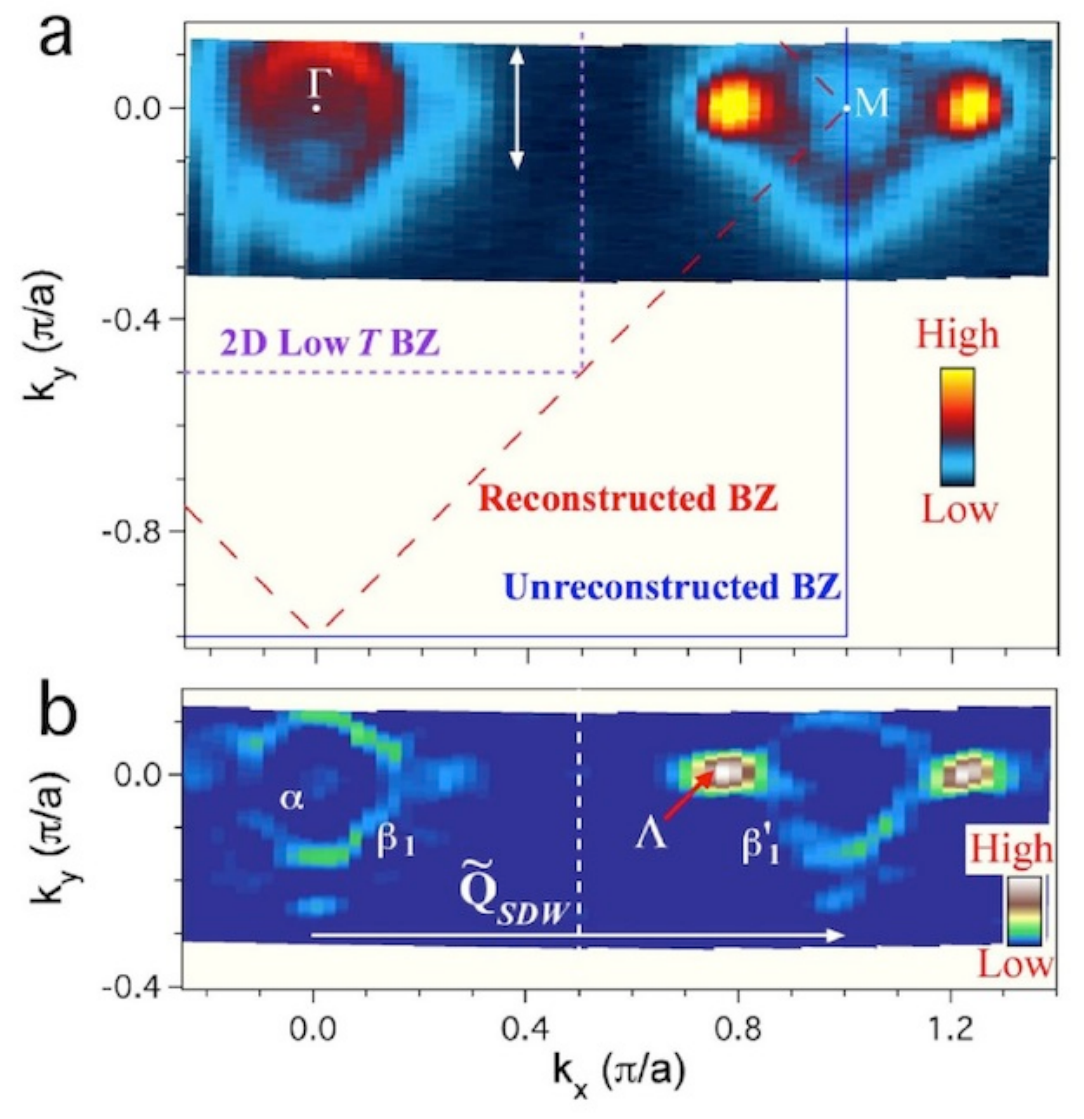}
\caption{\label{fig_parent} (Colour online) (a) FS mapping (25 K) obtained by integrating the photoemission intensity in a 20 meV window centered at $E_F$. The FS is described in terms of the unreconstructed BZ. The double arrow indicates light polarization. (b) Corresponding second derivative intensity plot. $\widetilde{Q}_{SDW}$ is the in-plane projection of the SDW wave vector. From Richard \emph{et al.}, Phys. Rev. Lett., \textbf{104}, 137001 (2010) \cite{RichardPRL2010}, copyright \copyright\xspace (2010) by the American Physical Society.} 
\end{center}
\end{figure}

Magnetic ordering in the parent compounds is accompanied by a reduction of the BZ and subsequent band folding. The relationship between the unreconstructed, reconstructed and 2D magnetic BZ in the 122 phase is described in Figure \ref{fig_parent}(a), which also shows the Fermi surface of BaFe$_2$As$_2$ well below the SDW transition $T_{{SDW}}=138$ K. The intensity mapping indicates Fermi surfaces of equivalent sizes around the $\Gamma$ and M points. The FS contours are better represented by the second derivative of the intensity map shown in Figure \ref{fig_parent}(b), which reveals almost identical patterns around $\Gamma$ and M, confirming band folding across the magnetic BZ boundaries. However, these patterns are quite unusual. In particular, strong photoemission intensity spots are observed near $\Lambda=(0, 0.75)$ and equivalent symmetry points, away from high symmetry points \cite{RichardPRL2010}. As temperature increases, these high intensity spots remain quite clear below $T_{{SDW}}$. Above that critical temperature, the spots are hard to identify, suggesting that they originate from the magnetic state. 

\begin{figure}[htbp]
\begin{center}
\includegraphics[width=8cm]{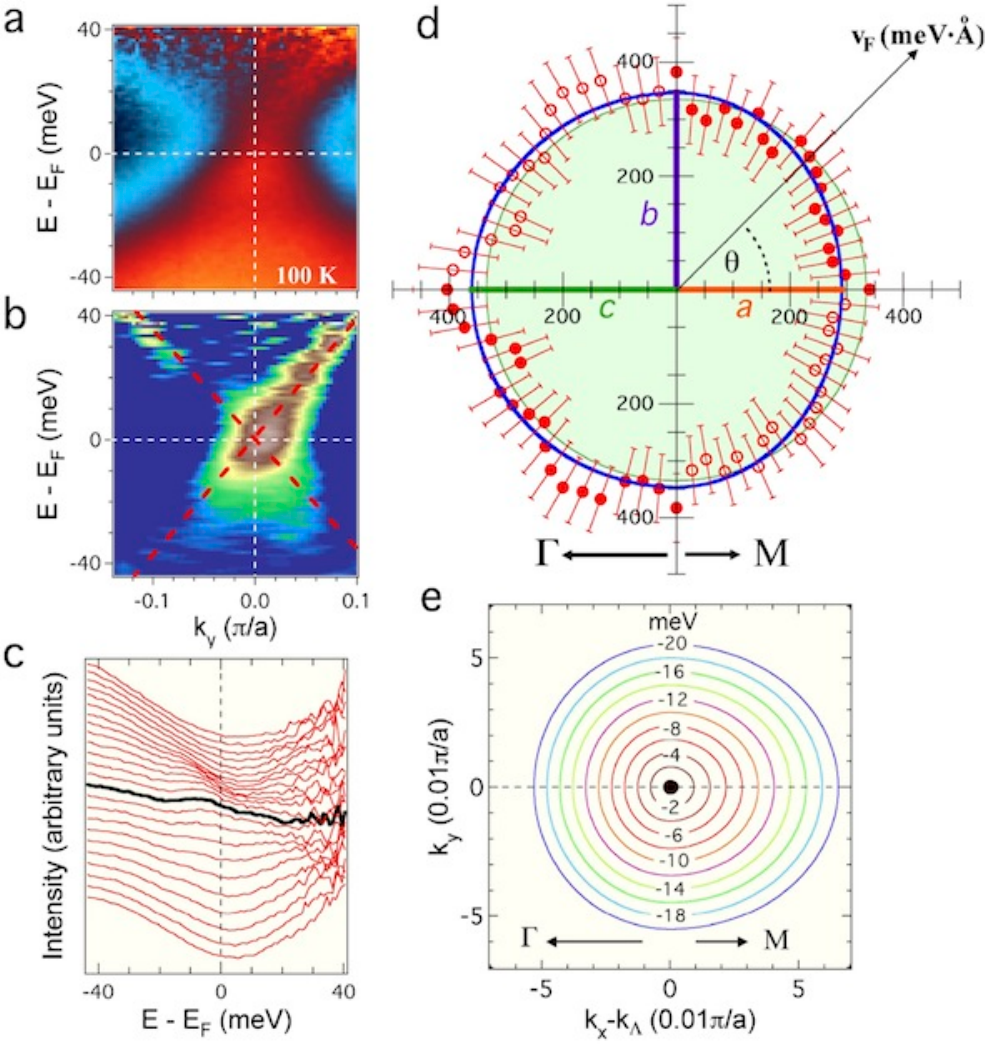}
\caption{\label{fig_cone} (Colour online) (a) ARPES intensity spectra of BaFe$_2$As$_2$ at the $\Lambda$ point ($\theta\sim 90^{\circ}$) recorded at 100 K, after division by the Fermi-Dirac function convoluted with the instrumental resolution function. (b) Corresponding second derivative intensity plot. Long dashed lines are guides for the eye. The corresponding EDCs are given in (c), where the bold EDC refers to the $\Lambda$ point. (d) Polar representation of $v_F$ around the $\Lambda$ point (25 K). Open and close circles represent data measured and data obtained by reflection with respect to the $\Gamma$-M symmetry line, respectively. The large filled circle represents the average value of $v_F$ while the thick line is a fit of the data to the two-ellipse model described in the text, with parameters $a$, $b$ and $c$. (e) Contour plot of the electronic dispersion below $E_F$ around the $\Lambda$ point, as calculated from our model. The small filled circle represents the FS associated with the cone. From Richard \emph{et al.}, Phys. Rev. Lett., \textbf{104}, 137001 (2010) \cite{RichardPRL2010}, copyright \copyright\xspace (2010) by the American Physical Society.} 
\end{center}
\end{figure}

Interestingly, the intensity spots at the $\Lambda$ point cannot emerge from a single band. A careful analysis shows that two bands cross the Fermi level around the $\Lambda$ point. Even more surprising is the almost conical band structure at that particular momentum point, which cannot be produced by simple band folding. Using the broadening of the Fermi cutoff at higher temperature, one can access states above $E_F$ within a range of a few $k_BT$'s. The spectra recorded at 100 K, which has been divided by the Fermi function convoluted by the energy resolution function, is displayed in Figure \ref{fig_cone}(a). Along with its second derivative intensity plot [Figure \ref{fig_cone}(b)], it reveals a X-like pattern rather than a simple $\Lambda$-like one. This strongly suggests the presence of a Dirac cone \cite{RichardPRL2010}, which is further confirmed by the absence of hybridization gap at the cone apex, as illustrated by the EDC's in Figure \ref{fig_cone}(c).  

As shown in Figure \ref{fig_cone}(d), the analysis of the low-temperature band structure of BaFe$_2$As$_2$ suggests that the Fermi velocity $v_F$ varies slightly around the Dirac cone, from 290 $\cdot$ \AA\xspace to 360 meV$\cdot$ \AA, with an average of 330 $\pm$ 60 meV$\cdot$\AA, which is compatible with the symmetry of the $\Lambda$ point \cite{RichardPRL2010}. The apex is found at  1 $\pm$ 5 meV above $E_F$, implying a very small hole Fermi surface pocket covering only around 10$^{-3}$\% of the first Brillouin zone, as suggested by Figure \ref{fig_cone}(e). Interestingly, bulk-sensitive quantum oscillation measurements led to similar results for the Fermi velocity \cite{Harrison}, as well as for the presence of small Fermi surface pockets \cite{Sebastian,Analytis}. 

\begin{figure}[htbp]
\begin{center}
\includegraphics[width=8cm]{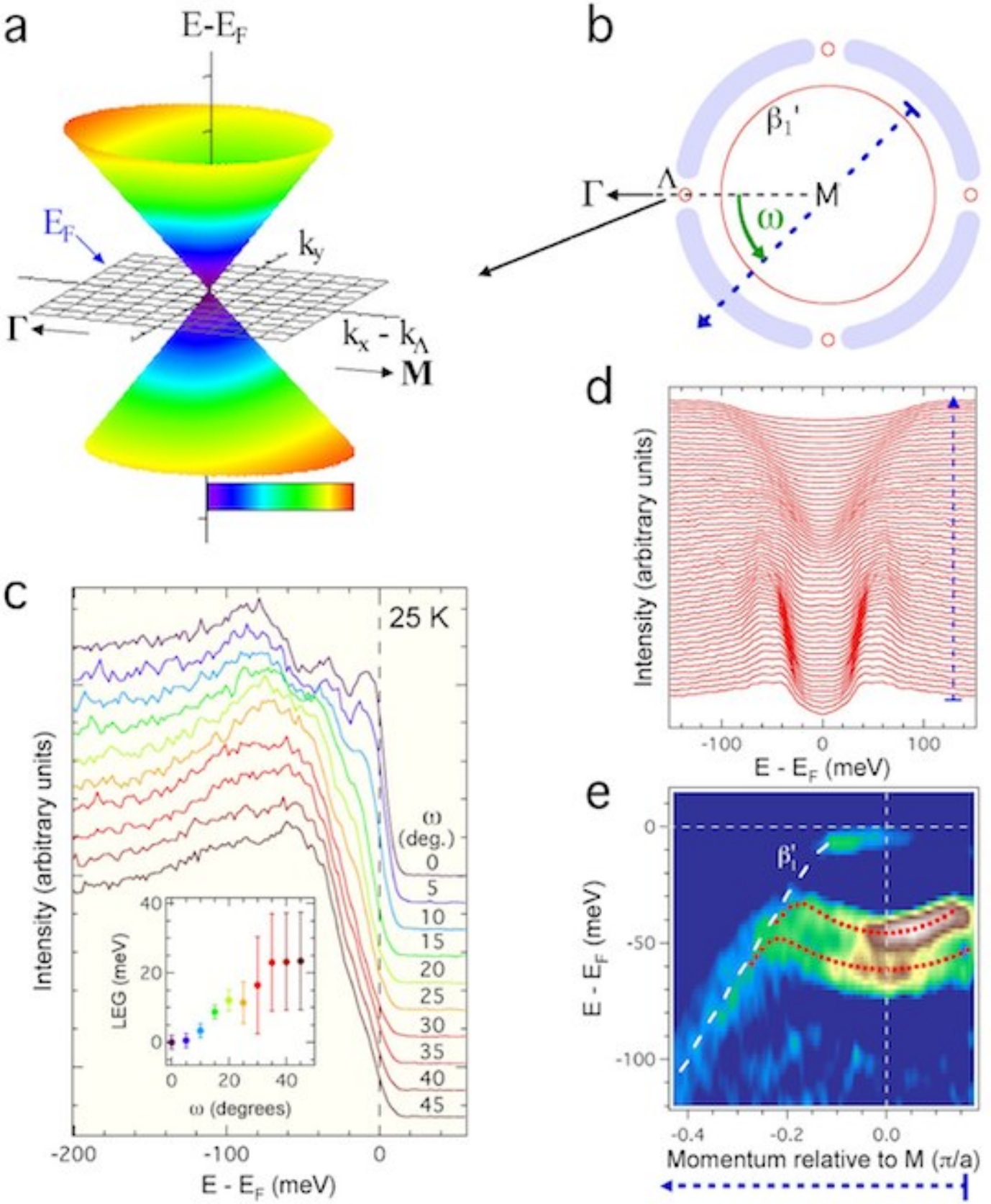}
\caption{\label{cone3D} (Colour online) (a) 3D representation of the Dirac cone at $\Lambda$. The colour scale indicates the distance from $\Lambda$. (b) Schematic FS around the M point. The folded $\alpha$ band, which is barely touching $E_F$, is not indicated. Shaded areas indicate gapped regions and the dashed arrow indicates the orientation of the ARPES cut associated with (d) and (e). (c) Minimum gap location of EDCs (25 K) as a function of the angle $\omega$ defined in (b). The inset shows the leading edge gap (LEG) as a function of $\omega$ after a 3.5 meV shift (see the text). (d) Symmetrized EDCs (25 K) along the cut indicated in (b). (e) Second derivative intensity plot (25 K) along the cut indicated in (b). The vertical dashed line indicates the M point. From Richard \emph{et al.}, Phys. Rev. Lett., \textbf{104}, 137001 (2010) \cite{RichardPRL2010}, copyright \copyright\xspace (2010) by the American Physical Society.} 
\end{center}
\end{figure}

The access to $v_F$ around the Dirac cone allows a 3D reconstruction of the Dirac cone, which is given in Figure \ref{cone3D}(a). Although such Dirac cones in the magnetic phase of these materials were not commonly expected, they were predicted by Ran \emph{et al.} \cite{Ran}, who argued that even in the presence of perfect nesting, the degeneracy of the band structure at the $\Gamma$ and M points must lead to nodes in the spin-density-wave gap function. Indeed, their calculations indicate the formation of Dirac cones at the SDW gap nodes. The SDW gap increases away from the Dirac cone, in agreement with ARPES measurements of the leading edge shift around the M point (as defined in Figure \ref{cone3D}(b)), which are shown in Figure \ref{cone3D}(c). Figures \ref{cone3D}(d) and (e) confirm that the electronic states are gapped away from the Dirac cone. Interestingly, ARPES measurements indicate that the electronlike bands have gaps of $\sim 30$ and $\sim 50$ meV. In a first approximation, considering that the Dirac cone apex is almost at the Fermi level, the corresponding full SDW gaps can be estimated at $\sim 60$ and $\sim 100$ meV, which is not far from the values of 45 and 110 meV determined from optical data \cite{WZ_HuPRL2008}. The presence of Dirac cones in the Fe-based superconductors goes much beyond superconductivity and connects several materials which appear completely different at the first sight \cite{Hassan_physics2010}. For example, time-reversal symmetry and $C_3$ crystal structure induce Dirac cones in graphene, a single layer of carbon atoms on a triangular lattice. Dirac fermions also emerge at the surface of the newly discovered topological insulators as a consequence of spin-orbit coupling. Finally, the cuprates, which have a Mott insulating parent compound, display a Dirac cone at the nodal point of their $d$-wave superconducting gap once they are doped with impurities. The massless dispersion common to all these materials is believed to play a major role in the next generation of electronic and spintronic devices. Not only Fe-based superconductors now join this category of materials, they constitute the first illustration by ARPES of an anisotropic Dirac cone dispersion. In addition, this discovery shows the importance that orbital characters may play in Fe-based superconductivity \cite{Hassan_physics2010}.

Recent DMFT calculations in the magnetic state of BaFe$_2$As$_2$ show a very nice consistency with experimental data \cite{Yin_ZP}. Unlike most ARPES experiments performed on twinned samples, the calculations exhibit the expected 2-fold symmetry rather than a 4-fold symmetry. To reveal the in-plane anisotropy of the band structure from ARPES, samples have to be untwined \emph{in situ} before the experiment \cite{M_YiPNAS2011,Y_KimPRB2011}. 

Contrary to optimally-doped \BKFAop, where only a small warping of the band structure is reported \cite{YM_Xu_NPhys2011}, a strong variation of the electronic band structure along the $k_z$ direction is found experimentally \cite{VilmercatiPRB2009, LiuPRL2009, FinkPRB2009, Malaeb_JPSJ2009, M_YiPRB2009_2,Brouet_PRB2009,Kondo_PRB2010,M_YiPNAS2011}. However, only the $\alpha$ band shows a significant $k_z$-dependence, with a periodicity of 4$\pi/c$ due to the periodicity of the primitive unit cell. While a 3D $\alpha$ pocket centered at $\Gamma$ has been reported for CaFe$_2$As$_2$ \cite{LiuPRL2009}, some reports on BaFe$_2$As$_2$ suggest that the $\alpha$ band crosses $E_F$ at every $k_z$ \cite{Malaeb_JPSJ2009}. In contrast to the $\alpha$ band, the bands forming the Dirac cone, as well as the electronlike bands at the M point, are only weakly sensitive to $k_z$. Interestingly, one report on CaFe$_2$As$_2$ clearly associates the $k_z$ variations to the SDW state \cite{LiuPRL2009}. Above $T_{{SDW}}$, the band structure does not show obvious modulations along $k_z$. 

The 11-chalcogenide system has the simplest structure of all Fe-based superconductors. Interestingly, its magnetic ground state is also different. Unlike BaFe$_2$As$_2$, the AF wave vector of FeTe points in the $\Gamma$-X direction. Consequently, the electronic band structure folds with respect to different magnetic zone boundaries below the ordering temperature. This effect has been observed directly by ARPES \cite{Y_XiaPRL2009,Y_Zhang_PRB2010}, where bands are folded to the X point, as expected. 

\subsection{Doping evolution of the electronic states in the 122-pnictide system}\label{section_doping}

\begin{figure}[htbp]
\begin{center}
\includegraphics[width=8cm]{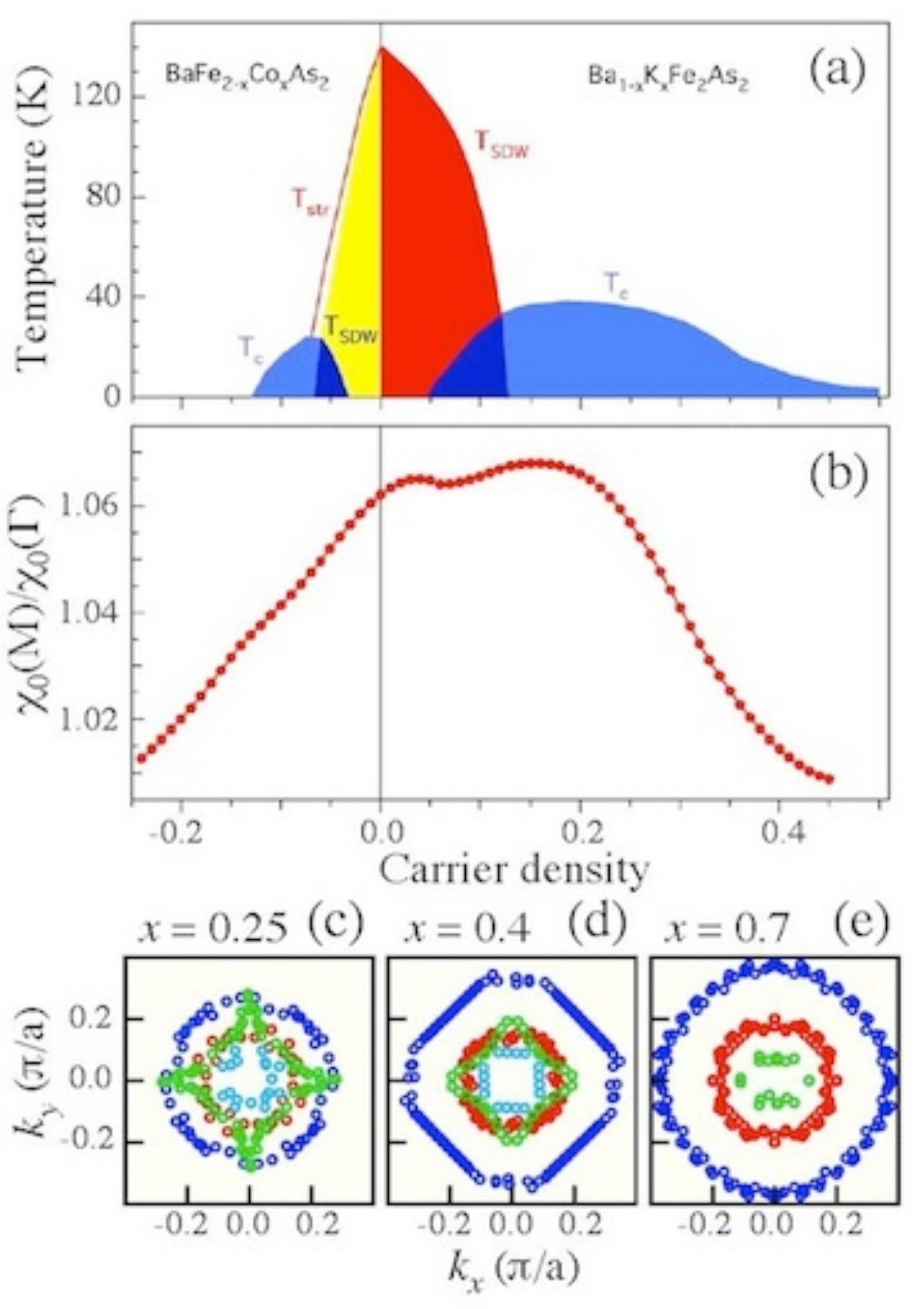}
\caption{\label{Phase_diagram} (Colour online) (a) Phase diagram of the hole- and electron-doped Ba-122 systems taken from Refs. \cite{Rotter_Angew2008} and \cite{PrattPRL2009}, respectively. $T_c$, $T_{SDW}$, and $T_{str}$ refer to the SC, the SDW, and the tetragonal to orthorombic structural transitions, respectively. (b) Doping dependence of the Lindhard function $\chi_0$ at the M point (quasi-nesting wave vector) normalized by its value at the zone center. The Lindhard function was obtained by using LDA calculations. (c)-(e) Experimentally determined $k_F$ points of the $\alpha$, $\beta$, and $\gamma/\delta$ bands [red (gray), blue (dark gray), and green (light gray) circles, respectively]. The $k_F$ points of the $\gamma/\delta$ FSs are shifted by $\vec{Q}=(-\pi,0)$. (a) and (b) are from Neupane \emph{et al.}, Phys. Rev. B, \textbf{83}, 094522 (2011) \cite{Neupane_PRB2011}, copyright \copyright\xspace (2011) by the American Physical Society.  (c)-(e) are from Nakayama \emph{et al.}, Phys. Rev. B, \textbf{83}, 020501(R) (2011) \cite{Nakayama_PRB2011}, copyright \copyright\xspace (2011) by the American Physical Society.} 
\end{center}
\end{figure}

The 122-pnictides have been the most studied among the Fe-based superconductors and we already have a rough overview of the phase diagram as seen by ARPES. The schematic phase diagram of these materials is illustrated in Figure \ref{Phase_diagram}(a). Both the electron-doped and the hole-doped sides show superconductivity and antiferromagnetism. A complication occurs in the electron-doped side, where the spin-density-wave transition does not coincide with the structural transition that accompanies the spin-density-wave transition in the hole-doped case. However, the main difference between the two sides of the phase diagram is the respective sizes of the antiferromagnetic and SC regions. The SC critical temperature reaches a maximum of about 37 K in the hole-doped case, which is obtained at a doping of about 0.2 doping hole per Fe, whereas it tops only around 25 K in the electron-doped materials, with an optimal doping that is roughly 0.08 electron dopant per Fe. Moreover, the SC dome is wide for hole doping and a SC tail survives up to the end of the phase diagram with KFe$_2$As$_2$. It is likely that the reason for such behaviour lies in the electronic band structure itself.

With the electronic carrier concentration varying, the Fermi surface of bulk states must evolve according to the Luttinger theorem, with the size of electronlike Fermi surface pockets increasing (decreasing) as we add electrons (holes), and the size of holelike Fermi surface pockets behaving exactly the other way. An illustration of the size evolution of the Fermi surface pockets in hole-doped Ba$_{1-x}$K$_{x}$Fe$_2$As$_2$ around optimal doping is given in Figures \ref{Phase_diagram}(c)-(e). For convenience, the M-centered electronlike FS pockets have been shifted to the $\Gamma$ point. As expected, the size of the $\alpha$ and $\beta$ holelike FS pockets increases with the hole carrier concentration increasing from underdoped Ba$_{0.75}$K$_{0.25}$Fe$_2$As$_2$ \cite{YM_Xu_UD} to overdoped Ba$_{0.3}$K$_{0.7}$Fe$_2$As$_2$ \cite{Nakayama_PRB2011} while the size of the electronlike pockets is reduced. Precise analysis of the algebraic FS area is consistent with the bulk concentration in each case.

To characterize the similarity between the size of the $\alpha$ Fermi surface and that of the M-centered electronlike Fermi surface pockets, here we extend the notion of nesting beyond its strict meaning. In 2D, if two circular electronlike and holelike pockets overlap completely under translation by the momentum $\vec{Q}$, the bare static spin susceptibility $\chi_0(\vec{q}, E=0)$ diverges logarithmically as $\vec{q}$ approaches $\vec{Q}$. This singularity defines perfect nesting with the important consequence that the random phase approximation (RPA) susceptibility diverges as well, \emph{i.e.} the system becomes unstable towards an SDW with wave vector $\vec{Q}$ for arbitrarily weak interactions. Although $\chi_0(\vec{q}, E=0)$ no longer diverges at $\vec{Q}$ if the size and shape of the Fermi surfaces start to deviate slightly from each other, the susceptibility can remained peaked at $\vec{Q}$, and thus moderate interactions can still drive the system into a SDW or CDW transition in the presence of fluctuations near the wave vector $\vec{Q}$. The precise behaviour of the susceptibility for what becomes natural to call \emph{near-nesting} or \emph{quasi-nesting} can be analytically understood in terms of a cutoff for the divergence arising from the deviation to perfect nesting \cite{Cvetkovic_EPL2009}. Since the Fermi surfaces are simply the constant energy contours of the energy dispersions at zero energy, the electronlike and holelike pockets can be even better nested by low-energy contours close to the Fermi level. This implies that quasi-nesting is even more robust for dynamical fluctuations, for which $\chi_0(\vec{q}, E)$ is very much enhanced, producing strong low-energy fluctuations with wave vector $\vec{Q}$, as indicated by the numerical renormalization group approach \cite{F_WangEPL2009}. In practice, we will say that the holelike and electronlike Fermi surfaces are quasi-nested if for large portions of the holelike Fermi surface, we can find a small wave vector $\delta\vec{q}_i$ and a small energy $\delta E_i$ so that the Fermi wave vector $\vec{k}_{F,i}$ on the holelike Fermi surface can be connected by the wave vector $\vec{Q}+\delta\vec{q}_i$ to one point on the $\delta E_i$ energy contour of the electronlike dispersion.

Within the extended notion of nesting described above, the $\alpha$ Fermi surface is well quasi-nested with the M-centered electronlike Fermi surface pockets in the optimally-doped system. Unavoidably, the quasi-nesting conditions must evolve while doping the system. With underdoping, the size of the $\beta$ Fermi surface shrinks but is still far to match the Fermi surface at the M point. In the meantime, the quasi-nesting of the M-centered electron Fermi surfaces and the $\alpha$ Fermi surface deteriorates only slightly, mainly along the $\Gamma-M$ direction. In the slightly overdoped region, the quasi-nesting conditions are more severely affected. Yet, each Fermi surface survives and antiferromagnetic scattering between $\Gamma$ and M is still expected. This statement is no longer true when doping the system even more with holes. 

\begin{figure}[htbp]
\begin{center}
\includegraphics[width=8cm]{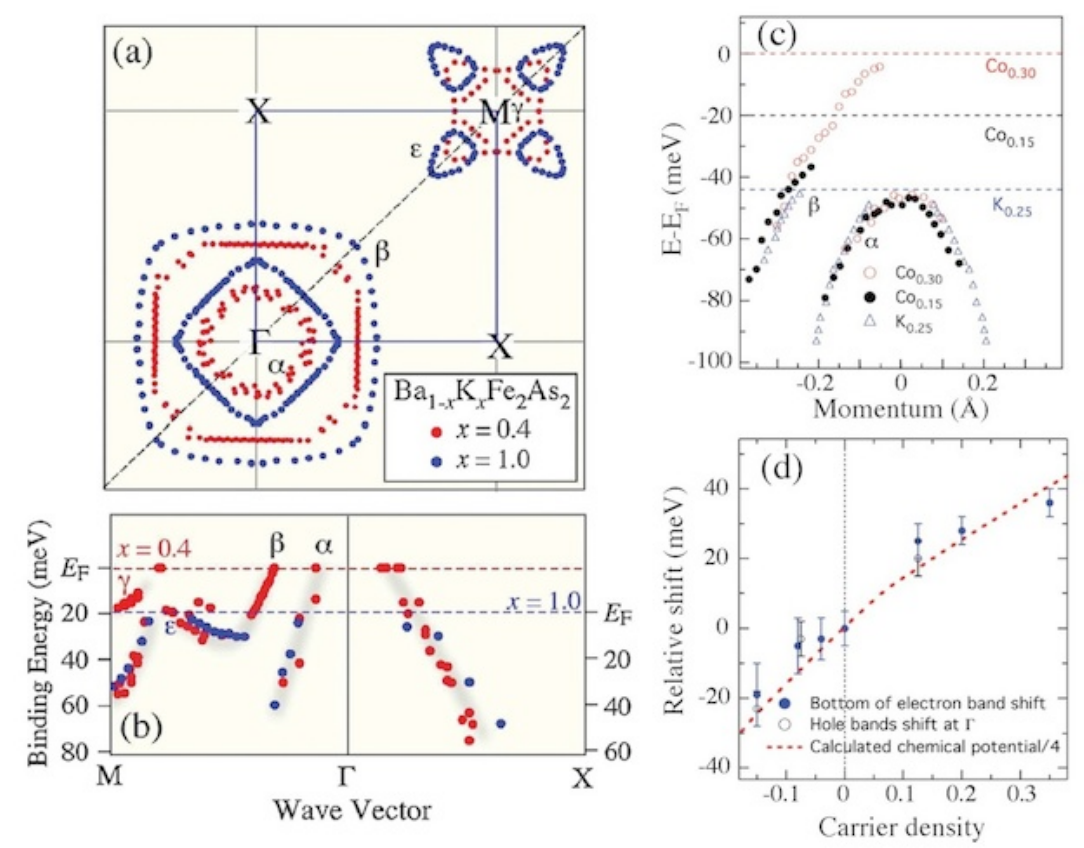}
\caption{\label{mu_shift} (Colour online) (a) Comparison (in the 2 Fe/unit cell notation) of experimentally determined $k_F$ points between overdoped KFe$_2$As$_2$ ($T_c=3$ K) and optimally doped Ba$_{0.6}$K$_{0.4}$Fe$_2$As$_2$ ($T_c=37$ K) \cite{Ding_EPL} (blue and red circles, respectively). The $k_F$ points are symmetrized by assuming a fourfold symmetry with respect to the  $\Gamma$ and M points. (b) Experimental band dispersion in the vicinity of $E_F$ for two high-symmetry lines determined by tracing the peak position of the ARPES spectra. The chemical potential of the KFe$_2$As$_2$ sample is shifted downward with respect to that of the Ba$_{0.6}$K$_{0.4}$Fe$_2$As$_2$ sample. (c) Fits of the $\alpha$ and $\beta$ bands for cuts passing through the $\Gamma$ point. The bands have been shifted to match the band slopes. (d) The blue dots are the valence band shifts shown in panel (c) and the red dash curve is the LDA calculated values of the chemical potential divided by a factor of 4. The black dots correspond to the relative shifts derived from panel (c), with the shift of the BaFe$_{1.7}$Co$_{0.3}$As$_2$ compound fixed arbitrarily on the renormalized LDA curve. (a) and (b) are from Sato \emph{et al.}, Phys. Rev. Lett., \textbf{103}, 047002 (2009) \cite{Sato_PRL2009}, copyright \copyright\xspace (2009) by the American Physical Society. (c) and (d) are from Neupane \emph{et al.}, Phys. Rev. B, \textbf{83}, 094522 (2011) \cite{Neupane_PRB2011}, copyright \copyright\xspace (2011) by the American Physical Society.} 
\end{center}
\end{figure}

Sato \emph{et al.} reported a detailed analysis of the band structure of fully K-substituted Ba$_{1-x}$K$_x$Fe$_2$As$_2$ \cite{Sato_PRL2009}. The main finding of this report is the total absence of electronlike Fermi surface at the M point in KFe$_2$As$_2$, which is equivalent to a suppression of the quasi-nesting condition between holelike and electronlike Fermi surface pockets. Instead, four small Fermi surfaces, elongated along the $\Gamma$-M direction, emerge away from M, as indicated in Figure \ref{mu_shift}(a). Interestingly, this dramatic change in the Fermi surface topology is not caused by a reconstruction of the band structure, but simply by a $\sim 20$ meV shift of the chemical potential as compared to the optimally-doped compound (see Figure \ref{mu_shift}(b)).     

The optimally electron-doped BaFe$_{1.85}$Co$_{0.15}$As$_2$ system ($T_c=25.5$ K) has first been studied by Terashima and co-workers \cite{Terashima_PNAS2009}. Due to the chemical potential shift that accompanies electron doping, the $\alpha$ Fermi surface pocket is totally absent in this material. Instead, only the $\beta$ Fermi surface survives around the $\Gamma$ point. Nevertheless, its size is significantly reduced. On the other hand, the M-centered electronlike Fermi surface pockets expand, as expected, and the size of the inner electronlike pocket at M becomes similar to the size of the $\beta$ pocket, suggesting that antiferromagnetic scattering can become important for this composition. As with its holelike counterpart, this is not the case when the system is extremely electron-doped. Sekiba \emph{et al.} demonstrated the absence of holelike Fermi surface pocket and thus the disappearance of quasi-nested Fermi surfaces in BaFe$_{1.7}$Co$_{0.3}$As$_2$, in which $T_c$ vanishes \cite{Sekiba_NJP2009}. Indeed, even the $\beta$ band sinks completely below $E_F$ at that high electron doping. In contrast, the M-centered electronlike Fermi surfaces become larger than in BaFe$_{1.85}$Co$_{0.15}$As$_2$, as required by the Luttinger theorem. We note that this change in the Fermi surface topology as compared to the optimally-doped system is induced by a chemical potential shift of about 20 meV due to electron-doping and that the overall band structure is preserved \cite{Sekiba_NJP2009}.  

Unlike the hole-doped side of the phase diagram, traces of what can be attributed to long-range antiferromagnetism survives at quite high electron doping in the underdoped regime, at least when using normalized doping $x/x_c$. Indeed, the bright spots attributed to the band-folding-induced Dirac cone in the parent compound \cite{RichardPRL2010} are observed until the onset of superconductivity is reached \cite{C_LiuNaturePhys2010}. For the electron-doped compounds though, the bright spots evolve into small holelike pockets \cite{C_LiuNaturePhys2010}. We note that their elongated shape is consistent with the anisotropy of the Dirac cone \cite{RichardPRL2010}. 

A critical observation in the 122-pnictides is that, besides the band folding that is present in the antiferromagnetic regime, the band structure of these materials does not vary significantly with doping. Instead, the electronic states near the Fermi level are tuned by a simple shift of the chemical potential with doping. Neupane \emph{et al.} tracked the position of the bottom of the M-centered electronlike bands as a function of doping on both side of the phase diagram \cite{Neupane_PRB2011}. Similarly, the authors also estimated the shift in the position relative to $E_F$ of the $\Gamma$-centered holelike bands by matching their electronic dispersions, as illustrated in Figure \ref{mu_shift}(c). Interestingly the shift of the electronic band structure at the $\Gamma$ and M points coincide, at least at the first order. More importantly, Figure \ref{mu_shift}(d) shows that the chemical potential shift determined by this method is in good agreement with the one estimated from LDA calculations after renormalization of the band width by a factor of 4 \cite{Neupane_PRB2011} to take into account the stronger renormalization observed for the electronic states very near $E_F$ \cite{Ding_J_Phys_Condens_Matter2011}. This conclusion has two important consequences. First, it confirms experimentally that the system is really doped through the substitution of Fe by Co, and that this substitution does not affect the effective masses of the system significantly, in contrast to speculations from a theoretical model that suggest that the Co substitution does not dope the system \cite{WadatiPRL2010}. Second, it suggests that renormalized LDA band calculations can be used to predict various electronic behaviours of these compounds, at least at the first order and away from long-range antiferromagnetic ordering. Starting from this assumption, Neupane \emph{et al.} calculated the doping evolution of the Lindhard function at the antiferromagnetic wave vector \cite{Neupane_PRB2011}. The result, displayed in Figure \ref{Phase_diagram}(b) shows a surprising similarity with the phase diagram, if we disregard the antiferromagnetic region for reasons mentioned above. For example, the maximum value of the Lindhard function is higher on the hole-doped side than in the electron-doped side, and the position of that maximum is also found at higher doping. Furthermore, the Lindhard function keeps a relatively high value for a wider range of doping in the hole-doped side. Overall, this surprising result suggests some correlation between superconductivity and antiferromagnetic scattering near-$E_F$ in the 122-pnictides. 

\subsection{Other materials}
\label{Other_materials}

Except for the 122-chalcogenides, which will be described at the end of this section, all Fe-based superconductors share the same fermiology for most part of their phase diagram: circular to slightly squarish holelike Fermi surface pockets centered at $\Gamma$ and a pattern formed by the hybridization of two ellipse electronlike Fermi surface pockets at the antiferromagnetic wave vector $M=(\pi,0)$. However, the quasi-nesting conditions between these two types of Fermi surfaces vary from one compound to another. For example, relatively good quasi-nesting is found in the electron-doped 111-pnictide NaFe$_{0.95}$Co$_{0.05}$As \cite{ZH_LiuPRB2011}. In contrast, the quasi-nesting conditions, as defined in Section \ref{section_doping}, are much weaker\footnote{We note that Borisenko \emph{et al.} use the word ``nesting" in reference \cite{BorisenkoPRL2010} in a much stricter sense than the concept of quasi-nesting defined in Section \ref{section_doping}, and thus claim that there is no nesting at all in LiFeAs.} in SC LiFeAs ($T_c=18$ K) \cite{BorisenkoPRL2010}, even though slightly inelastic antiferromagnetic scattering near the Fermi surface remains possible . It is also interesting to point out that all predicted holelike bands are observed in the 111-pnictides near the $\Gamma$ point \cite{C_HePRL2010,ZH_LiuPRB2011,BorisenkoPRL2010} with much less ambiguity than in the 122-pnictides. As with these later materials, the band structure of the 111-pnictides is also renormalized significantly near $E_F$ compared to band structure calculations. A renormalization factor of 3 has been reported for LiFeAs \cite{BorisenkoPRL2010} while He \emph{et al.} report renormalization by factors varying between 4.3 and 5.4 on different bands in the parent compound NaFeAs \cite{C_HePRL2010}. 

Despite an obvious surface problem induced by the absence of appropriate cleaving plane that results in excess hole doping at the surface layer, ARPES results on the 1111-pnictides \cite{DH_LuNature2008, Kondo_PRL2008, C_LiuPRB2010, HY_LiuPRL2010} also indicate essentially the same fermiology as their pnictide cousins. In addition, experimental data indicate a renormalization of the band structure in LaOFeP of 2.2 \cite{DH_LuNature2008}, which is similar to the other pnictides. Unlike for the other Fe-based superconductors though, no variation of the band structure is observed by ARPES when changing the photon energy \cite{C_LiuPRB2010}, which is easily understood since the state probed is a real surface state induced by the cleaving problem for this material. 

\begin{figure}[htbp]
\begin{center}
\includegraphics[width=8cm]{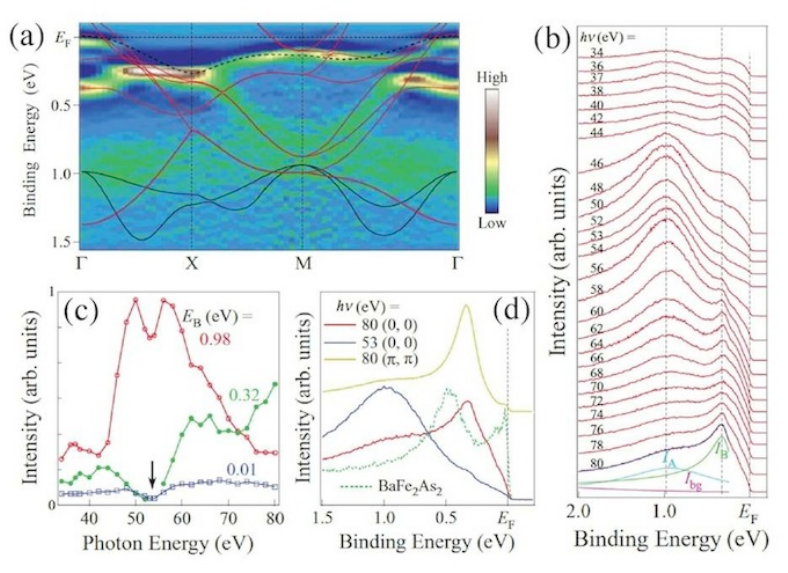}
\caption{\label{V_paper} (Colour online) (a) Second derivative plot of along the $\Gamma$-X-M-$\Gamma$ high symmetry lines of Sr$_2$VFeAsO$_3$. LDA+U bands are also plotted for comparison. The Fe $3d$ bands (red/gray lines) are renormalized by a factor of 1.6, whereas the V $3d$ bands (black lines) are not. The Fe 3$d_{xy}$ band (black dashed line) is renormalized by a factor of 3.3 to reproduce the experimental band near $E_F$. (b) Valence band (VB) at the BZ center measured at different photon energies (34-80 eV). All the spectra are normalized by the photon flux. The fitting curves for the spectrum at $h\nu=80$ eV are also shown. (c) Photon energy dependence of the intensities of the two peaks obtained from the fitting curves and the spectral intensity at $E_B=0.01$ eV. (d) Direct comparison of the VB measured at $h\nu=53$ and 80 eV, along with the VB of BaFe$_2$As$_2$ ($h\nu=80$ eV). From Qian \emph{et al.}, Phys. Rev. B, \textbf{83}, 140513 (2011) \cite{Qian_PRB2011}, copyright \copyright\xspace (2011) by the American Physical Society.} 
\end{center}
\end{figure}

Among all pnictides, a special attention must be devoted to Sr$_2$VFeAsO$_3$, for which there exists, to our knowledge, only one ARPES report \cite{Qian_PRB2011}. In addition to Fe3$d$ states, most LDA band calculations predicted the presence of V3$d$ states at the Fermi level that would completely modify the fermiology of this system \cite{KW_LeeEPL2010,MazinPRB2010_21311,Shein_JSNM2009,G_WangPRB2009}. In fact, one of these papers mentioned that if the Fermi surface is derived only from Fe states, it would have a topology comparable to that of the other pnictides \cite{MazinPRB2010_21311}. This is precisely what is found experimentally \cite{Qian_PRB2011}. Using samples smaller than $0.2\times 0.2$ mm$^2$, Qian \emph{et al.} determined that the Fermi surface of this system is composed by a small and a large $\Gamma$-centered holelike Fermi surfaces (called $\alpha$ and $\beta$, respectively) and by electronlike Fermi surfaces with elongated ellipse shape at the M point. The electron counting from the algebraic Fermi surface area leads to a carrier concentration compatible with the Luttinger theorem. Comparing their results with LDA+U calculations, the authors found a good agreement with the overall Fe3$d$ band structure when considering a normalization factor of 1.6, as shown in Figure \ref{V_paper}(a). However, the states closer to the Fermi level fit well the calculation only for a renormalization factor of about 3.3 \cite{Qian_PRB2011}. The main discrepancy between the experimental results and the other LDA calculations relates to the position of the V states. From their resonance profile as a function of incident photon energy, illustrated in Figures \ref{V_paper}(b)-(d), the V3$d$ states are found around 1 eV below $E_F$, a result that is reproduced only when incorporating an effective Hubbard energy U of 6 eV in the LDA+U calculations \cite{Qian_PRB2011}. It is interesting to note that with V$3d$ electrons in a Mott state and Fe$3d$ electrons at the Fermi level in the absence of Fe-V hybridization, Sr$_2$VFeAsO$_3$ can be viewed as a perfect orbital selective Mott transition \cite{AnasimovEPJB2002,LiebschPRL2003} system, such as in Ca$_{1.8}$Sr$_{0.2}$RuO$_4$, which has also been characterized by ARPES \cite{NeupanePRL2009}.

The Fe-chalcogenides have been less studied than the pnictides by ARPES, even the 11-chalcogenides that have been synthesized in the early days of Fe-based superconductivity. Nevertheless, existing data of doped 11-chalcogenides show spectral intensity at the $\Gamma$ and M$(\pi,0)$ points \cite{Nakayama_PRL2010, Tamai_PRL2010,Miao2011}. Although up to three bands can be identified at $\Gamma$, the situation remains more nebulous at the M point, making a careful electron counting analysis difficult. As with pnictides, some high energy features can be well approximated by LDA calculations renormalized by a factor around 2 \cite{Nakayama_PRL2010}. However, additional renormalization much stronger than in the pnictides is found near $E_F$ \cite{Nakayama_PRL2010, Tamai_PRL2010}, with renormalization factors as large as 20 \cite{Tamai_PRL2010}. 

\begin{figure}[htbp]
\begin{center}
\includegraphics[width=8cm]{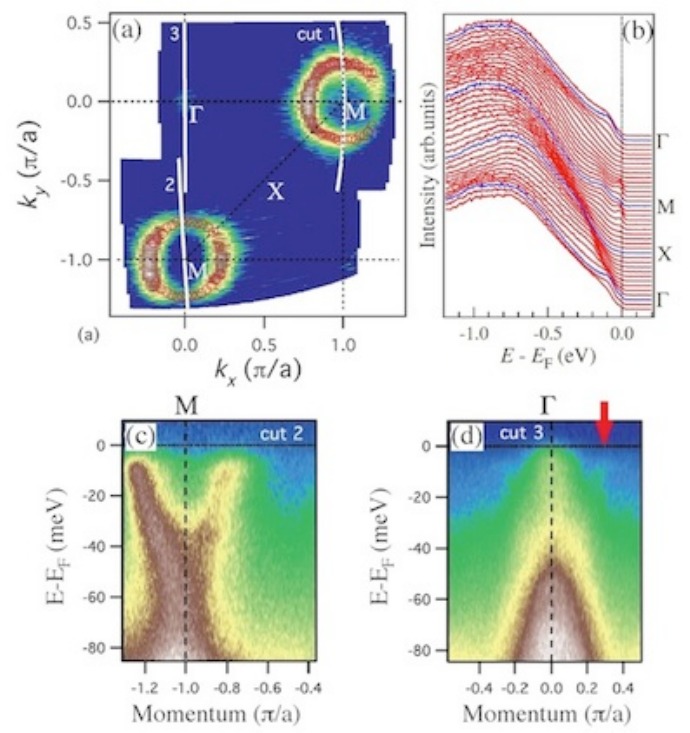}
\caption{\label{new122_structure} (Colour online) (a) Momentum-resolved photoemission intensity mapping of Tl$_{0.63}$K$_{0.37}$Fe$_{1.78}$Se$_2$ recorded in the normal state (35 K) and integrated over a 10 meV window centered at $E_F$. The small red circles indicate the FS obtained from the momentum distribution curve (MDC) peak position at $E_F$. (b) EDCs along several high symmetry directions in K$_{0.8}$Fe$_{1.7}$Se$_2$ recorded with the He I resonance line ($h\nu=21.218$ eV). Blue (dark gray) curves correspond to high symmetry points. (c) ARPES intensity plot along a cut passing through the M point (cut1 from (a)). (d) ARPES intensity plot along a cut passing through the $\Gamma$ point (cut3 from (a)). (a), (c) and (d) are from Wang \emph{et al.}, Europhy. Lett., \textbf{93}, 57001 (2011) \cite{XP_WangEPL2011}, copyright \copyright\xspace (2011) by the European Physical Society. (b) is from Qian \emph{et al.}, Phys. Rev. Lett., \textbf{106}, 187001 (2011) \cite{Qian_PRL2011}, copyright \copyright\xspace (2011) by the American Physical Society.} 
\end{center}
\end{figure}

Recently, the landscape of Fe-based superconductivity changed abruptly with the discovery of high-$T_c$ ($T_c=29$ K) in heavily-electron-doped K$_x$Fe$_2$Se$_2$ \cite{JG_Guo_PRB2010} and related materials. The input of ARPES to characterize and explain the electronic properties of these materials has been very helpful. Unlike all the other Fe-based superconductors, this system does not show any holelike Fermi surface \cite{Qian_PRL2011, XP_WangEPL2011, Y_Zhang_NatureMat2011, D_MouPRL2011}, as illustrated in Figure \ref{new122_structure}(a). In contrast, a large electronlike Fermi surface is observed at the M point. Although the zone center is free of holelike Fermi surfaces, there is evidence for additional electronlike pockets whose origin are not quite clear at the moment. A small 3D electronlike pocket centered at the Z point has been reported in a synchrotron-based study on A$_x$Fe$_2$Se$_2$ (A = K, Cs) \cite{Y_Zhang_NatureMat2011}. Traces of these pockets have also been detected in (Tl,K)Fe$_{1.78}$Se$_2$ \cite{XP_WangEPL2011} and Tl$_{0.58}$Rb$_{0.42}$Fe$_2$Se$_2$ \cite{D_MouPRL2011} using He discharge lamps. These laboratory-based experiments also reveal an additional electronlike pocket with a size more or less similar to the one centered at the M point \cite{XP_WangEPL2011, D_MouPRL2011}, as we can see from the comparison of Figures \ref{new122_structure}(c) and (d).  

As with the other Fe-based superconductors, the 122-chalcogenides have their overall band width renormalized by a factor of about 2.5 \cite{Qian_PRL2011}. What is different though is the energy distribution of the spectral weight, which is closer to the one observed in the cuprates. While the spectral weight of the Fe3$d$ electrons is much larger in the other Fe-based superconductors, the near-$E_F$ bands in K$_{0.8}$Fe$_{1.7}$Se$_2$ have much weaker intensity. On the other hand, the intensity of the valence band is more important in this materials. In fact, Figure \ref{new122_structure}(b) shows that the weight below 2 eV is mainly concentrated in a broad peak located around 800 meV below $E_F$, which does not show significant dispersion. That peak, which has been attributed to the incoherent part of the Fe3$d$ states \cite{Qian_PRL2011}, shows an important energy shift at a temperature $T^*$ corresponding to a bump in the resistivity curves \cite{XP_WangEPL2011}.   

\section{Superconducting gap}
\label{section_gap}

The order parameter describing the SC phase transition is characterized by a complex function represented by the momentum distribution of the gap developing at $E_F$ below the SC critical temperature $T_c$. The amplitude and phase of the order parameter are directly determined by the electronic structure and by the mechanism responsible for the pairing of charge carriers. Although ARPES is not directly sensitive to the phase, it allows precise measurement of the momentum dependence of the SC gap size, which serves as a powerful tool to validate or invalidate theoretical models. The previous statement is particularly true for multi-band systems, for which the interpretation of transport measurements, which may be momentum-sensitive but not momentum-resolved, becomes difficult or even impossible. 

\begin{figure}[htbp]
\begin{center}
\includegraphics[width=8cm]{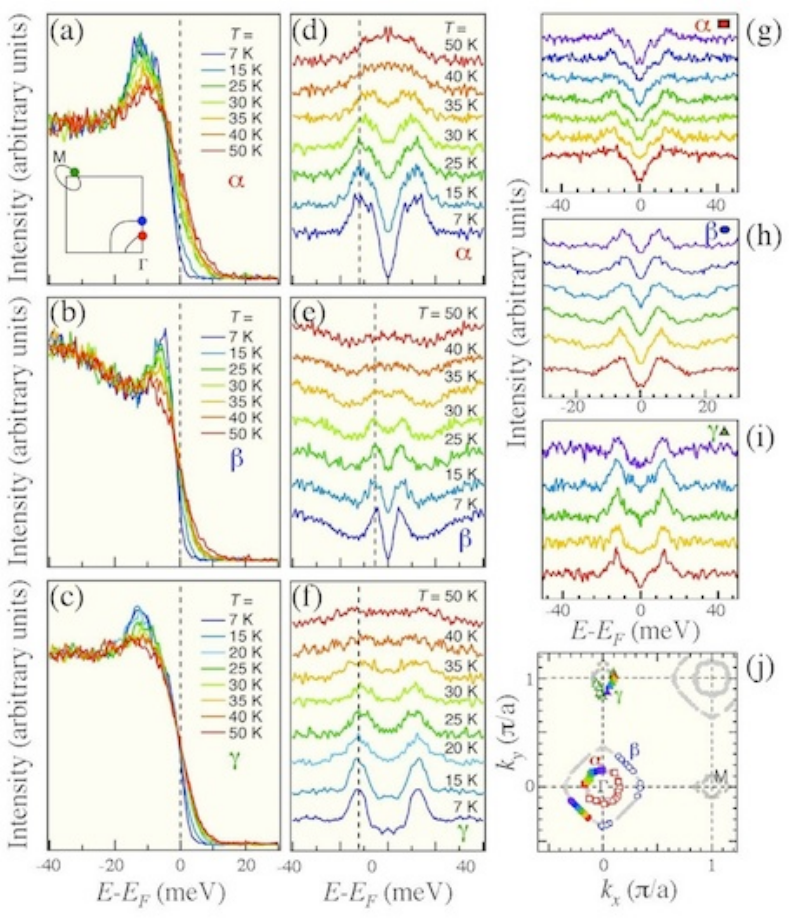}
\caption{\label{Gap_OP} (Colour online) Superconducting gap of \BKFAop. (a)-(c) Temperature dependence of the EDCs on the $\alpha$, $\beta$ and $\delta$ Fermi surfaces, respectively. The $k_F$ position of the EDCs is given on the schematic FS (in the 2 Fe/unit cell representation) displayed in the inset of panel (a). (d)-(f) Corresponding symmetrized EDCs. (g)-(i) Momentum dependence of symmetrized EDCs at 15 K along the $\alpha$, $\beta$ and $\delta$ Fermi surfaces, respectively. The $k_F$ position are given on the FS displayed in panel (j) in the 1 Fe/unit cell representation. Squares, circles and triangles correspond to the $k_F$ positions on the $\alpha$, $\beta$ and $\delta$ Fermi surfaces, respectively. Filled and empty symbols refer to real and symmetrized data, respectively. Adapted from Ding \emph{et al.}, Europhy. Lett., \textbf{83}, 47001 (2008) \cite{Ding_EPL}, copyright \copyright\xspace (2008) by the European Physical Society.} 
\end{center}
\end{figure}

\begin{figure}[htbp]
\begin{center}
\includegraphics[width=8cm]{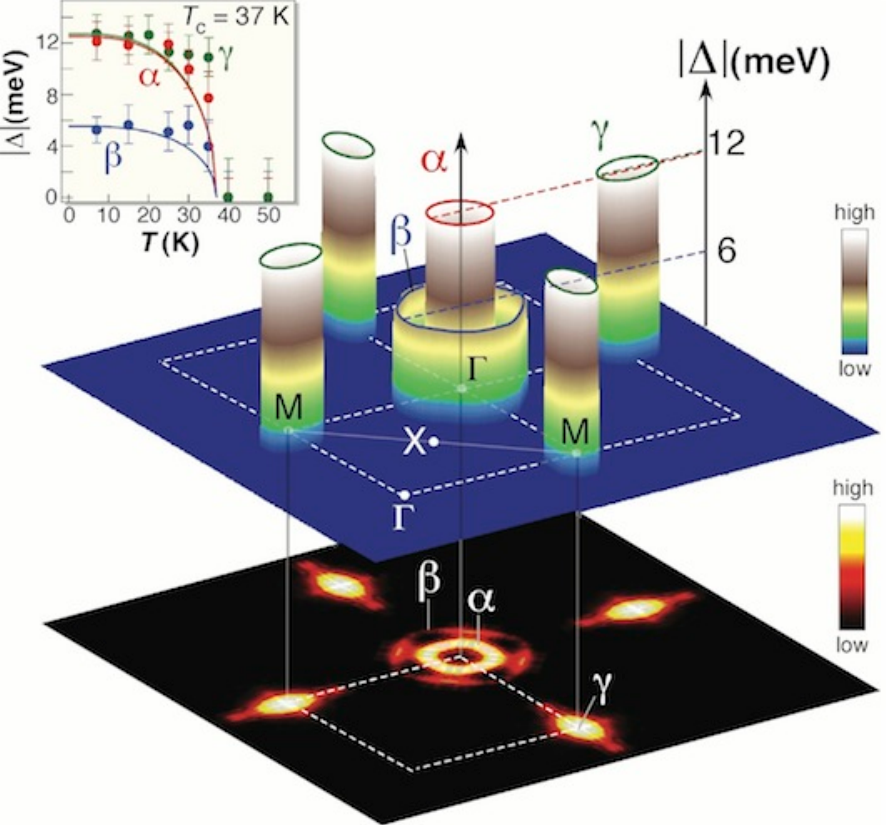}
\caption{\label{EPL_final} (Colour online) Three-dimensional plot of the SC gap size ($\Delta$) measured at 15 K on three FS sheets (shown at the bottom as an intensity plot) and their temperature evolutions (inset). From Ding \emph{et al.}, Europhy. Lett., \textbf{83}, 47001 (2008) \cite{Ding_EPL}, copyright \copyright\xspace (2008) by the European Physical Society.} 
\end{center}
\end{figure}

The first ARPES determinations of the SC gap in Fe-based superconductors were obtained on optimally-doped \BKFAops \cite{Ding_EPL,L_Zhao}. Figures \ref{EPL_final}(a)-(c) show the temperature evolution of EDCs from Ding \emph{et al.} recorded with the He I$\alpha$ resonance line on the $\alpha$, $\beta$ and $\gamma$ Fermi surfaces, respectively \cite{Ding_EPL}. As temperature decreases below $T_c$, a coherence peak develops on each FS, indicating the opening of a superconducting gap, which is better visualized from the corresponding symmetrized EDCs displayed in Figures \ref{EPL_final}(d)-(f). The full superconducting gap size $2\Delta$ is given by the distance between the two coherence peaks in the symmetrized EDCs. The superconducting gap is also characterized by its momentum dependence on each FS, which are shown in Figures \ref{EPL_final}(g)-(i). The $k_F$ points where these data were collected are indicated in Figure \ref{EPL_final}(j).  

Figure \ref{EPL_final} summarizes the data from Ding \emph{et al.} \cite{Ding_EPL}. Unlike the SC gap in the cuprates, these experiments show clearly the absence of nodes along all the Fermi surface pockets, ruling out any order parameter with a $d$-wave symmetry. The invariance of the coherence peak positions in the symmetrized EDCs shown in Figures \ref{EPL_final}(g)-(i), as well as further refined measurements with the He I$\alpha$ line \cite{Nakayama_EPL2009}, indicate that the gaps are indeed isotropic. Interestingly though, the gap size is not unique and varies from band to band. More precisely, the gap size is around 12 meV along the quasi-nested FSs, yielding to $2\Delta/k_BT_c$ ratios of about 7.5, while the gap size along the non-quasinested $\beta$ FS pocket drops to 6 meV. We note that although the SC gap seems to disappear above $T_c$, the temperature evolution of the SC gap size does not follow the BCS function and shows a much steeper drop while approaching $T_c$. Isotropic gaps have been reported not only in the 122-pnictides \cite{Ding_EPL,L_Zhao}, but also in the 111-pnictide \cite{ZH_LiuPRB2011,BorisenkoPRL2010}, the 1111-pnictide \cite{Kondo_PRL2008} and the 11-chalcogenide \cite{Nakayama_PRL2010} systems. 

To explain these intriguing results, two main models have been proposed for the pairing mechanism: the \emph{FS quasi-nesting model} and the \emph{local AF exchange pairing model}. These two models are radically different. Essentially, the basic question to discriminate one from the other is: Does pairing occur due to interactions in the momentum-space or in the real-space? Historically, the former model gained popularity very quickly, partly due to ARPES results. We thus describe it first, along with the reasons for its initial success, among which are the evidence for the presence of interband scattering in the pnictides and the weakening of the quasi-nesting conditions accompanying the disappearance of superconductivity at very high doping in these materials. Paradoxically, recent ARPES measurements on the 122-chalcogenides are main indicators announcing the dusk of that model to describe the Fe-based superconductors. Nowadays, the local AF exchange pairing model described afterwards appears as a better prospect in claiming for universality of the pairing mechanism in the Fe-based superconductors. Yet, the community is still far from ready to unanimously award laurel wreath to any model candidate. 

\subsection{Quasi-nesting model}
\label{section_QN}

\begin{figure}[htbp]
\begin{center}
\includegraphics[width=8cm]{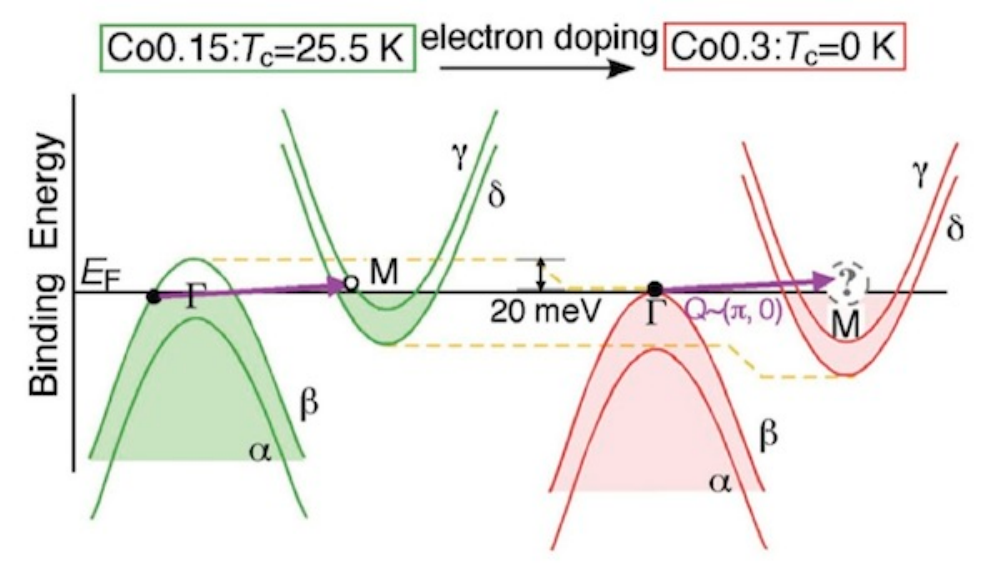}
\caption{\label{fig_nesting} (Colour online) Comparison of energy bands between BaFe$_{1.85}$Co$_{0.15}$As$_2$ and BaFe$_{1.7}$Co$_{0.3}$As$_2$ samples. The interband scattering is dramatically suppressed in the non-SC BaFe$_{1.7}$Co$_{0.3}$As$_2$ compound since the holelike $\alpha$ and $\beta$ bands at the $\Gamma$ point are basically occupied. From Sekiba \emph{et al.}, New J. Phys., \textbf{11}, 025020 (2009) \cite{Sekiba_NJP2009}, copyright \copyright\xspace (2009) by IOP Publishing and Deutsche Physikalische Gesellschaft.} 
\end{center}
\end{figure}

Until the recent discovery of superconductivity in the chalcogenide version of the 122 phase, the quasi-nesting model was the most popular way to account for superconductivity in Fe-based superconductors. The idea is indeed very simple and illustrated in Figure \ref{fig_nesting}: when a holelike and an electronlike FS pockets have similar but not exact size and shape, the system can avoid long-range ordered CDW or SDW states but yet maintain important scattering induced by short-range fluctuations through the ``quasi-nesting" wave vector (in this case the antiferromagnetic wave vector) that connects these two FS pockets. Despite the absence of perfect nesting, a peak of enhanced magnetic susceptibility is still expected at the quasi-nesting wave vector, in which case the kinetic process by which a zero momentum pair formed on one FS is scattered onto another one can be enhanced, whereby increasing the pairing amplitude \cite{Ding_EPL, MazinPhysicaC2009, Graser_NJP2009}. 

\begin{figure}[htbp]
\begin{center}
\includegraphics[width=8cm]{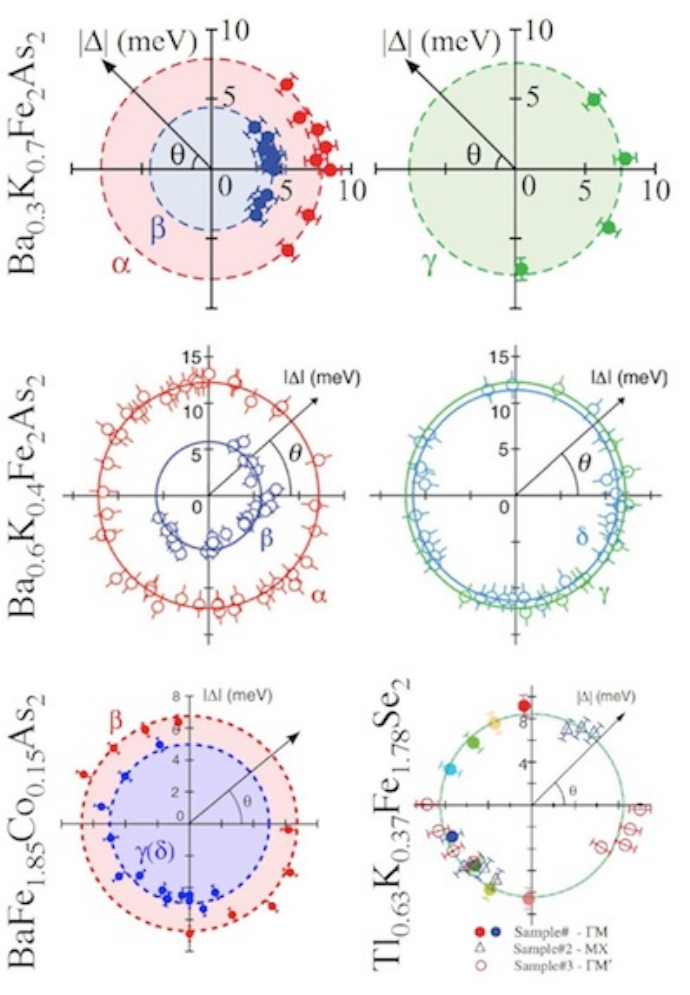}
\caption{\label{polar} (Colour online) Polar representation of the SC gap amplitude for various bands and materials. The data on Ba$_{0.3}$K$_{0.7}$Fe$_2$As$_2$ are from Nakayama \emph{et al.}, Phys. Rev. B, \textbf{83}, 020501(R) (2011) \cite{Nakayama_PRB2011}, copyright \copyright\xspace (2011) by the American Physical Society. The data on Ba$_{0.6}$K$_{0.4}$Fe$_2$As$_2$ are from Nakayama \emph{et al.}, Europhy. Lett., \textbf{85}, 67002 (2009) \cite{Nakayama_EPL2009}, copyright \copyright\xspace (2009) by the European Physical Society. The data on BaFe$_{1.85}$Co$_{0.15}$As$_2$ are from Terashima \emph{et al.}, Proc. Natl. Acad. Sci. USA, \textbf{106}, 7330 (2009) \cite{Terashima_PNAS2009}, copyright \copyright\xspace (2009) by the National Academy of Sciences of the United States of America. The data on Tl$_{0.63}$K$_{0.37}$Fe$_2$Se$_2$ are from Wang \emph{et al.}, Europhy. Lett., \textbf{93}, 57001 (2011) \cite{XP_WangEPL2011}, copyright \copyright\xspace (2011) by the European Physical Society.} 
\end{center}
\end{figure}

In practice, the quasi-nesting model can be tested by checking the evolution of the SC gap size as a function of the quasi-nesting conditions. The latter are easily tuned with doping since the doping evolution of holelike and electronlike FS pockets with carrier injection have opposite trends. The most drastic effect is observed when extra-doping the system until one type of FS pocket, either electronlike or holelike, disappears. As mentioned previously, the purely K-substituted 122 compound (KFe$_2$As$_2$) is extremely over-hole-doped and does not exhibit any electronlike FS \cite{Sato_PRL2009}. In agreement with the quasi-nesting model, its $T_c$ drops to a few Kelvins only. Unfortunately, the gap size becomes too small to be extracted. Similarly, the disappearance of $\Gamma$-centered holelike FS pockets in heavily electron-doped BaFe$_{1.7}$Co$_{0.3}$As$_2$ is accompanied by a total suppression of superconductivity \cite{Sekiba_NJP2009}. Another experimental fact consistent with the quasi-nesting model in the 122 system is the switch of the pairing strength on the $\beta$ band from ``weak coupling" to ``strong coupling" as we compare the optimally-hole-doped and optimally-electron-doped compounds (see Figure \ref{polar}). While the $\beta$ FS is quite larger than the M-centered electronlike FS pockets in the hole-doped compound, its size becomes compatible with the FS pockets at the M point in the optimally-electron-doped system, thus promoting AF interband scattering \cite{Terashima_PNAS2009}. 

\begin{figure}[htbp]
\begin{center}
\includegraphics[width=8cm]{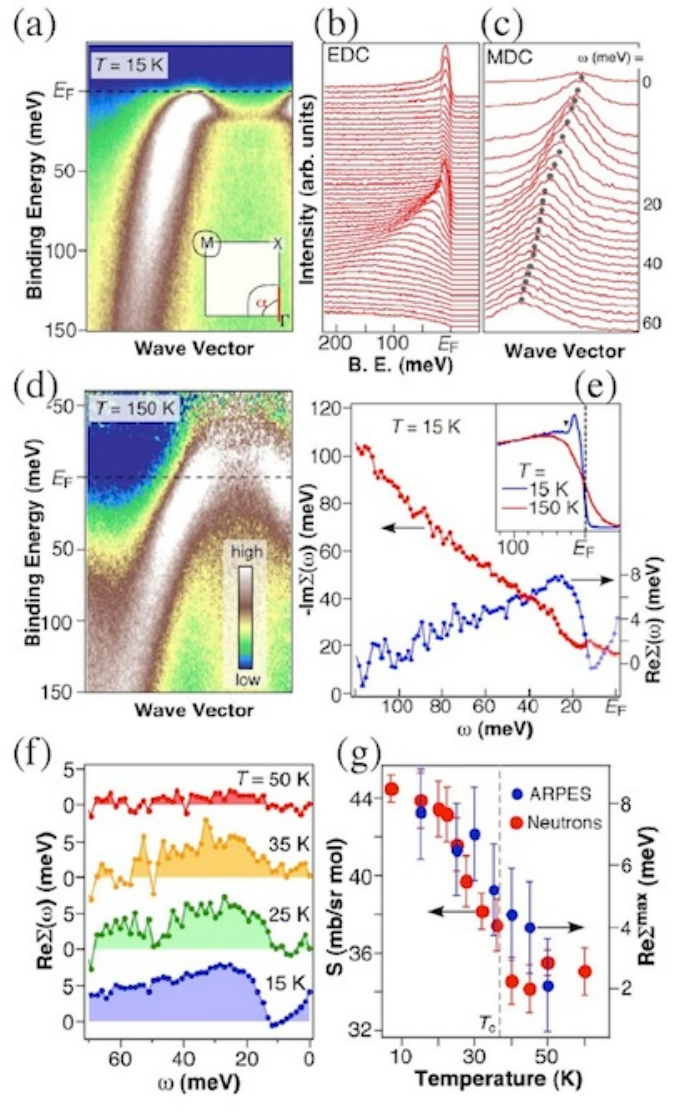}
\caption{\label{fig_kink} (Colour online) (a) ARPES intensity plot in the SC state (15 K) along a cut crossing the $\alpha$ band. The inset shows the schematic FS of Ba$_{0.6}$K$_{0.4}$Fe$_2$As$_2$ with the location of the cut (red) in the 2 Fe/unit cell notation. (b) Corresponding EDCs. (c) Corresponding MDCs, in the 0-60 meV binding energy range. Grey dots indicate the maximum position of the peaks. (d) ARPES intensity plot at 150 K divided by a Fermi-Dirac function, recorded along the same cut as in (a). (e) Real and imaginary parts of $\Sigma(\omega)$. Fade colours are used for binding energies smaller than 17 meV since $\Sigma(\omega)$ in this range is complicated by particle-hole mixing due to superconductivity. The inset compares the partial DOS along the cuts measured at 15 K (blue) and 150 K (red), respectively. (f) Temperature dependence of the real part of the self-energy referred from the MDC fit dispersion at 150 K. (g) Maximum value of the real part of the self-energy (blue) plotted as a function of temperature. The ARPES results are compared to the neutron scattering intensity of the 14 meV spin resonance (red) located at the AF wave vector \cite{Christianson}. The dashed line indicates the critical temperature. From From Richard \emph{et al.}, Phys. Rev. Lett., \textbf{102}, 047003 (2009) \cite{RichardPRL2009}, copyright \copyright\xspace (2009) by the American Physical Society.} 
\end{center}
\end{figure}

Whatever it is or not the cause for Fe-based superconductivity, ARPES provides sufficient evidence for interband scattering with the quasi-nesting wave vector. A first evidence is the smaller EDC linewidth of the unnested $\beta$ band in \BKFAops as compared to the quasi-nested bands \cite{Ding_EPL}. It indicates that the lifetime of the quasi-particles associated with the $\beta$ band is longer due the absence of interband scattering. A second evidence is the observation of an anomaly, or \emph{kink}, in the dispersion of bands associated with quasi-nested FS pockets \cite{RichardPRL2009}. At 15 K, this anomaly is found around 25 meV, as can been seen from Figures \ref{fig_kink}(a)-(c). To confirm this result, a self-energy analysis was performed (see Figure \ref{fig_kink}(e)) by approximating the bare band dispersion by the dispersion determined at 150 K, way above the critical temperature. This choice of bare band approximation, justified by the absence of anomaly at 150 K , is critical since the Fe-based superconductors are multi-band systems. In contrast, the cuprates are single-band systems studied not far from half filling, where the band dispersion near $E_F$ can be reasonably well approximated by a straight line. By dividing out the ARPES intensity plot recorded at 150 K by the Fermi-Dirac function, one can reveal the top of the $\alpha$ band about 25 meV above $E_F$, as illustrated in Figure \ref{fig_kink}(d), indicating that the band dispersion in the energy region of the kink is naturally curved and that a straight line cannot be used to approximate the bare band dispersion. In the SC state, both the imaginary and real parts of the self-energy deviate around 25 meV from the quite linear behaviour observed at higher energy. 

The coupling of an Einstein mode of energy $\Omega$ is expected to give rise to an anomaly at $\Omega+\Delta$ \cite{Norman_PRB1998}, where the SC gap $\Delta$ is well known for each band. Subtracting the 12 meV energy corresponding to the gap on the $\alpha$ FS from the kink energy yields $\Omega=13\pm 2$ meV. While an anomaly is also found in the quasi-nested $\gamma$ band, the $\beta$ band shows an anomaly neither at 25 meV nor at 18 meV ($\Delta_{\beta}+\Omega=6+13$ meV). This band-selectivity of the anomaly is in apparent contradiction with a conventional electron-phonon coupling and point towards an electronic origin \cite{RichardPRL2009}. Interestingly, the temperature dependence of the kink anomaly, shown in Figure \ref{fig_kink}(f), coincides with the temperature dependence of a spin resonance mode observed by inelastic neutron scattering at 14 meV \cite{Christianson}, as well illustrated in Figure \ref{fig_kink}(g). As with the ARPES kink, the spin resonance, detected at the antiferromagnetic wave vector, disappears above $T_c$, suggesting that the two phenomena are related.  

Similarly to the cuprates, antiferromagnetic scattering increases while under-doping Fe-based superconductors \cite{Ning_PRL2010}. This affects not only the width of the SC quasi-particle peaks, but also their weight. While the SC quasi-particle peak associated with the unnested $\beta$ band in Ba$_{0.75}$K$_{0.25}$Fe$_2$As$_2$ ($T_c=26$ K) remains sharp and coherent, the peaks associated with the nested $\alpha$ and $\gamma/\delta$ bands lose their integrity compared to the optimally-doped material \cite{YM_Xu_UD}. Indeed, this dichotomy between the behaviour of SC quasi-particle peaks on the unnested and quasi-nested FS pockets upon under-doping is quite similar to the one observed in cuprates between the nodal and antinodal regions. Unlike the quasi-particle peak in the nodal region, the quasi-particle peak at the antinode is quite suppressed.  
  
At least for the 122-pnictides, which has been more deeply studied, the doping evolution of the quasi-nesting conditions is in qualitative agreement with the variation of $T_c$. A good illustration of this statement is provided by the doping dependence of the renormalized Lindhard function at the antiferromagnetic wave vector given in Figure \ref{mu_shift}(b), which shows a strong electron-hole asymmetry reflecting the asymmetric band structure with respect to electron- or hole-doping \cite{Neupane_PRB2011}. While the Lindhard function keeps a high value for a wide hole-doping range before starting to decrease, it decreases monotonically on the electron-doped side, albeit for a small shoulder around x = 0.24. Although these calculations performed with the non-magnetic LDA band structure (renormalized by a factor of 4) are not suitable for the antiferromagnetic region near 0 doping, it qualitatively reproduces the size, height and shape of the SC dome away from 0 doping. 

Fair agreement with the quasi-nesting model is also obtained for electron-doped NaFe$_{0.95}$Co$_{0.05}$As ($T_c=18$ K), for which Liu \emph{et al.} reported almost identical SC gaps (6.8 \emph{vs} 6.5 meV) on the $\Gamma$-centered $\alpha^{\prime}$ FS pocket and on the M-centered electron FS pockets, which all have similar size \cite{ZH_LiuPRB2011}. This gap size leads to a $2\Delta/k_BT_c\sim 8$ ratio, indicating that the system is in a strong coupling regime. Using the leading edge shift rather than the SC coherent peak position to identify the gap size, Borisenko \emph{et al.} found a $2\Delta/k_BT_c$ ratio of about 3.5 in LiFeAs \cite{BorisenkoPRL2010}, which is more consistent with the BCS regime. Even though the leading edge method is necessarily an underestimation of the SC gap size, at least by the half-width at half maximum of the SC coherent peak, LiFeAs shows smaller gap size than NaFe$_{0.95}$Co$_{0.05}$As. Interestingly, the quasi-nesting conditions are also poorer in LiFeAs, which has one very large and one very small holelike $\Gamma$-centered FS pockets, along with intermediate size M-centered electronlike FS pockets. This reduction of both $T_c$ and the $2\Delta/k_BT_c$ ratio is thus compatible with the quasi-nesting model. 

As explained above, LDA calculations predicted the failure of the quasi-nesting model to explain $T_c=37$ K superconductivity in Sr$_2$VFeAsO$_3$ \cite{KW_LeeEPL2010,MazinPRB2010_21311,Shein_JSNM2009,G_WangPRB2009}. The arguments were mainly related to the presence of V states at $E_F$ in the calculated band structure. Although SC gaps were never measured by ARPES due to the small size of the samples, the experimental Fermi surface is quite similar to that of the other pnictides and thus consistent with the quasi-nesting scenario \cite{Qian_PRB2011}. A more intriguing case is that of the 11-chalcogenides. In contrast to the pnictides, the magnetic ground state of their parent compound is not described by the $(\pi,0)$ wave vector, but rather by a wave vector pointing in the $\Gamma-X$ direction. Yet, the FS topology of SC 11-chalcogenides is quite similar to that of the pnictides, and ARPES results by Nakayama \emph{et al.} suggest that the SC gap of the holelike band at $\Gamma$ is also isotropic and in the strong coupling regime \cite{Nakayama_PRL2010}. Unfortunately, the low spectral intensity at the M point did not allow the determination of the SC gap there until very recently \cite{Miao2011}. In any case, these surprising similarities between SC pnictides and 11-chalcogenides indicate the importance of the $(\pi,0)$ scattering, even though it does not necessarily proves the validity of the quasi-nesting model in explaining high-$T_c$ superconductivity in these materials. In fact, some neutron scattering experiments suggest that AF scattering at $(\pi,0)$ is important in this system and even evolves into a resonance for doped samples \cite{IikuboJPSJ2009,YM_QiuPRL2009,TJ_LiuNMAT2010}.

\subsection{Local antiferromagnetic exchange pairing model}
\label{section_local}

\begin{figure}[htbp]
\begin{center}
\includegraphics[width=8cm]{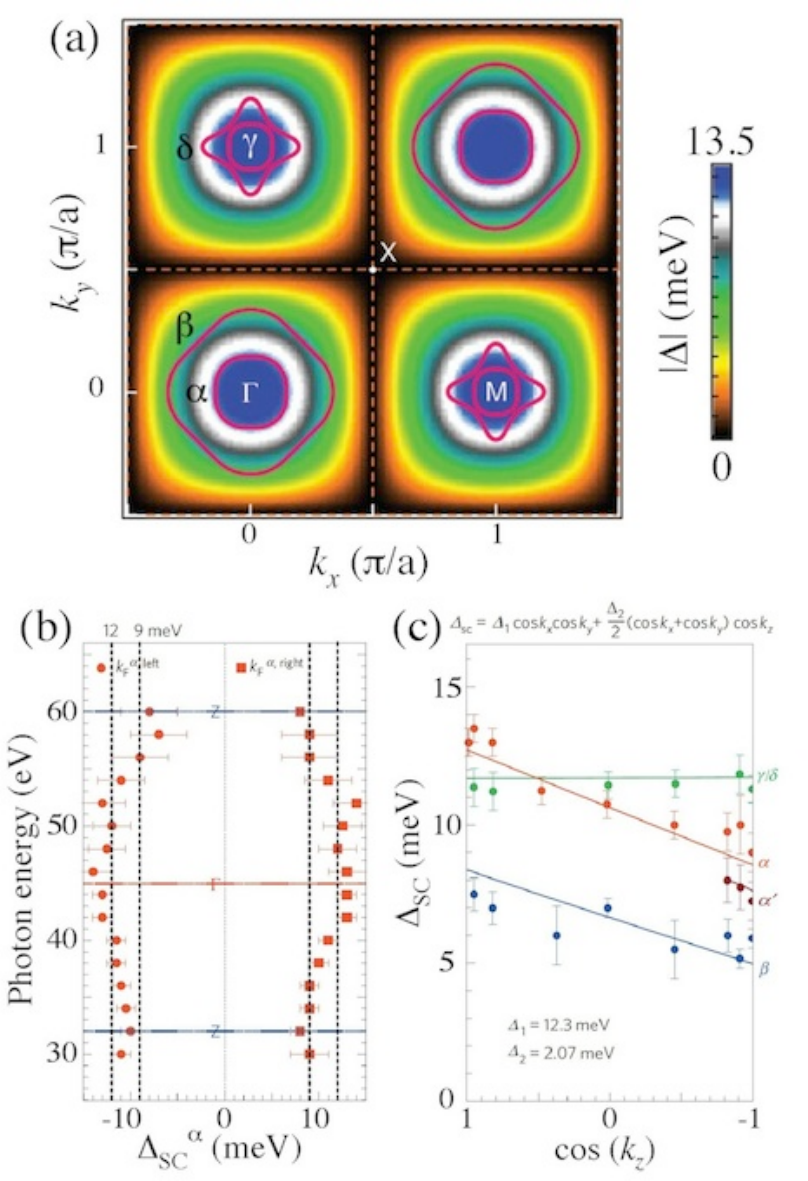}
\caption{\label{J_model} (Colour online) (a) Theoretical SC gap value $|\Delta (k) =  \Delta_0|\cos k_x \cos k_y|$ with $\Delta_0=13.5$ meV as a function of the two-dimensional wave vector \cite{SeoPRL2008,KorshunovPRB2008}. (b) Extracted values of the SC gap (defined as the half value of peak-to-peak positions in the symmetrized EDCs) on the $\alpha$ FS at different photon energies. The dots (squares) are obtained from the left (right) side of $k_F$ on the $\alpha$ FS. (c) The SC gap values on the $\alpha$ FS (red dots), $\beta$ FS (blue dots), $\gamma/\delta$ FS (green dots) and $\alpha$' FS (brown dots) as functions of the 3D gap function $|\Delta(k_x,k_y,k_z)|=|\Delta_1\cos k_x\cos k_y+(\Delta/2)(\cos k_x+\cos k_y)\cos k_z|$, with $\Delta_1 = 12.3$ meV and $\Delta_2=2.07$ meV, to fit all the SC gaps. The error bars are standard deviations of the measured SC gaps. Panel (a) is Nakayama \emph{et al.}, Europhy. Lett., \textbf{85}, 67002 (2009) \cite{Nakayama_EPL2009}, copyright \copyright\xspace (2009) by the European Physical Society. (b) and (c) are from Xu \emph{et al.}, Nature Phys., \textbf{7}, 198 (2011) \cite{YM_Xu_NPhys2011}, copyright \copyright\xspace (2011) by Macmillan Publishers Ltd.} 
\end{center}
\end{figure}

Enthusiasm for the quasi-nesting scenario in the pnictides was just too strong in the early days to allow the local pairing model to really take off. Nevertheless, Wray \emph{et al.} \cite{Wray_PRB2008} and Nakayama \emph{et al.} \cite{Nakayama_EPL2009} pointed out that the SC gap measured by ARPES in the 122-pnictides is qualitatively consistent with an order parameter that takes the form $\Delta_0\cos(k_x)\cos(k_y)$. Interestingly, such a formula can be derived from a picture where the pairing occurs in a short distance in real space, for example by considering antiferromagnetic exchange between nearest ($J_1$) and second-nearest ($J_2$) neighbors. Figure \ref{J_model}(a) displays the detailed comparison between this formula and the observed gap values obtained by Nakayama \emph{et al.} \cite{Nakayama_EPL2009}. Apart from small deviations, the same formula applies also to the 111-pnictide NaFe$_{0.95}$Co$_{0.05}$As, where similar $\Delta_0$ values of 6.8 and 6.5 meV are found for the $\Gamma$-centered holelike FS pockets and M-centered electronlike FS pockets, respectively \cite{ZH_LiuPRB2011}. 

The presence of a cosine oscillation along the $k_z$ direction of the gap value associated with the $\alpha$ band, well illustrated by Figure \ref{J_model}(b) justifies modifications of the $\Delta=\Delta_0\cos(k_x)\cos(k_y)$ formula. Xu \emph{et al.} \cite{YM_Xu_NPhys2011} started from the simple formula 

\begin{equation}
\Delta_{3D}(k_x,k_y,k_z)=\Delta_{2D}(k_x,k_y)(1+\eta\cos(k_z))
\end{equation}

\noindent which is a generalization to layered superconductors of the BCS expression for a superconductor with an isotropic in-plane gap function and where $\eta$ is a measure of the interlayer coupling strength. With specific considerations applying to the pnictides, they then derived the generalized $s$-wave gap function \cite{YM_Xu_NPhys2011}:

\begin{eqnarray}
\Delta=\Delta_1\cos(k_x)\cos(k_y)+\nonumber\\
(\Delta_2/2)(\cos(k_x)+\cos(k_y))\cos(k_z)+\Delta_3(\cos(k_x)\cos(k_y))\cos(k_z)
\end{eqnarray}

\noindent where the gap parameters $\Delta_1$, $\Delta_2$ and $\Delta_3$ can be orbital-dependent or band-dependent in the most general case. Neglecting the last term which is experimentally much smaller than the $\Delta_2$ term due to the vanishingly small $\eta$ of the M-centered electronlike FS pockets, the experimental data obtained on optimally-doped Ba$_{0.6}$K$_{0.4}$Fe$_2$As$_2$ can be well fitted with $\Delta_1=12.3$ meV and $\Delta_2=2.07$ meV \cite{YM_Xu_NPhys2011}. Figure \ref{J_model}(c) shows the fit of SC gap data on all the Fermi surfaces of Ba$_{0.6}$K$_{0.4}$Fe$_2$As$_2$ using a single gap function. Interestingly, the $\Delta_1/\Delta_2=5.9$ ratio coincide almost with the $J_{ab}/J_c=6$ ratio of the in-plane and out-of-plane magnetic coupling constants determined from spin-wave dispersions obtained by neutron scattering experiments on the parent compound \cite{J_Zhao_PRL2008}.

The validity of the gap function given above has been checked for over-doped Ba$_{0.3}$K$_{0.7}$Fe$_2$As$_2$ ($T_c=22$ K) with the He I$\alpha$ resonance line (21.218 eV). According to the $k_z$ dispersion of hole-doped 122-pnictides \cite{YM_Xu_NPhys2011}, the corresponding photon energy is close to the $\Gamma$ point, and thus $\cos(k_z)\simeq 1$. The experimental results yield $\Delta_1=8.3$ meV and $\Delta_2=0.7$ meV \cite{Nakayama_PRB2011}. It is worth nothing that within this model the electronlike and the holelike FS pockets may carry different gaps even if their size is the same since the term $\cos(k_x)+\cos(k_y)$ is 2 and 0 for the $\Gamma$ and M points, respectively.

Interestingly, the 3$^{rd}$ nearest neighbor antiferromagnetic exchange coupling $J_3$ in the $J_1$-$J_2$-$J_3$ model is very strong in the chalcogenides and must be taken into account \cite{LipscombePRL2011}. As a consequence, the formula $|\Delta (k) = \Delta_2 |\cos k_x \cos k_y|$ does not apply and must be replaced by $|\Delta (k)| = |\Delta_2 \cos k_x \cos k_y- \Delta_3(\cos 2k_x -\cos 2k_y)/2|$. The later fits pretty well the data of FeTe$_{0.55}$Se$_{0.45}$ ($T_c = 14$ K) and the 4.2 meV and 2.6 meV gaps found on the M-centered $\gamma$ FS and $\Gamma$-centered $\alpha$ FS assuming $\Delta_2 = 3.55$ meV and $\Delta_3 = 0.95$ meV \cite{Miao2011}. A noticeable feature is the stronger gap at the M point than at the $\Gamma$ point, which is not the case for the pnictides and that comes out naturally from the $J_1-J_2-J_3$ model. It also worth mentioning that the $\Delta_2/\Delta_3$ ratio is almost the same as the $J_2/J_3$ ratio determined experimentally from neutron scattering experiments \cite{LipscombePRL2011}. 

The revival of the local pairing picture is mainly due to the recent discovery of superconductivity at  high-temperature in over-electron-doped 122-chalcogenides. As mentioned in Section \ref{Other_materials}, these systems are completely free of holelike FS pockets at the $\Gamma$ point and thus prevent $(\pi,0)$ scattering in the electron-hole channel. Unless this new class of materials is the host of a different pairing mechanism, which is quite unlikely due to the structural and electronic similarities between the chalcogenides and the pnictides, the local pairing picture seems more robust as a universal model to explain superconductivity in the Fe-based superconductors. 

The SC gap of the 122-chalcogenides has been measured for (Tl,K)Fe$_{1.78}$Se$_2$ \cite{XP_WangEPL2011}, A$_x$Fe$_2$Se$_2$ \cite{Y_Zhang_NatureMat2011} and (Tl$_{0.58}$Rb$_{0.42}$)Fe$_{1.78}$Se$_2$ \cite{D_MouPRL2011}. All data suggest an isotropic SC around the M-centered electronlike FS pocket. In Tl$_{0.63}$K$_{0.37}$Fe$_2$Se$_2$ ($T_c=29$ K), a 8.5 meV gap is found, as indicated in Figure \ref{polar}, which leads to $2\Delta/k_BT_c\sim 7$, clearly in the strong coupling limit \cite{XP_WangEPL2011}. SC gaps are also reported for electronlike FSs centered at the $\Gamma$ point. Hence, Wang \emph{et al.} reported a 8 meV for the SC gap on the large $\Gamma$-centered electronlike FS pocket in Tl$_{0.63}$K$_{0.37}$Fe$_2$Se$_2$ \cite{XP_WangEPL2011}. Zhang \emph{et al.} reported a 7 meV on the small 3D electronlike pocket centered at the Z point in K$_{0.8}$Fe$_2$Se$_2$, as compared to 10.3 meV at the M point \cite{Y_Zhang_NatureMat2011}. Finally, Mou \emph{et al.} recorded a 12 meV gap on the large $\Gamma$-centered electronlike FS pocket in Tl$_{0.58}$Rb$_{0.42}$Fe$_{1.72}$Se$_2$ ($T_c=32$ K), in contrast to the 15 meV SC gap found at the M point \cite{D_MouPRL2011}. 
 
Following the recent developments on the SC gap of the 11-chalcogenides and the 122-chalcogenides as determined from ARPES measurements, the local antiferromagnetic exchange pairing model has been further developed to emphasize not only on the local exchange couplings, but also on the fermiology of the different materials \cite{Hu_universal}. More precisely, it is argued that high-temperature superconductivity occurs when the Fermi surface topology matches the form factor of the pairing symmetry favored by the local antiferromagnetic exchange interactions. The result extends to cuprates as well.

\subsection{Temperature dependence of the superconducting gap}

It  is has been noticed in an early ARPES report on the 122-pnictides that the temperature dependence of the SC gap was not the same as in the BCS superconductors \cite{Ding_EPL}. Unfortunately, the $\alpha$ band shows a ``shoulder" in the SC state that makes a precise characterization difficult. In contrast, the 111-pnictides constitutes an ideal system to investigate this issue since no shoulder is observed and the natural cleaved surfaces are expected to be non-polar and free of reconstruction. Liu \emph{et al.} performed such a study on electron-doped NaFe$_{0.95}$Co$_{0.05}$As \cite{ZH_LiuPRB2011}. From their data analysis, it is possible to see that the SC gap size $\Delta$, given by half the distance between the two SC coherent peaks, does not change much with temperature increasing. Instead, the gap is filling up as temperature increases. A more quantitative approach consists in fitting the experimental EDCs using the self-energy function $\Sigma(k,\omega)$ suggested by Norman \emph{et al.} for quasiparticles in the SC state \cite{Norman_PRB1998}: 

\begin{equation}
\Sigma(k,\omega)=-i\Gamma+\frac{\Delta^2}{(\omega+i0^+)+\epsilon(k)}
\end{equation}

\noindent where $\Gamma$ is the single-particle scattering rate, which is here assumed to be independent of $\omega$. Assuming a polynomial background, this function shows a good agreement with the experiments at all temperatures until the SC peak vanishes. As with the scattering rate and the SC gap size, the normalized coherent area $Z_A$ defined as $Z_A=\int A_{coh}(k,\omega)d\omega/\int A(k,\omega)d\omega$ as in earlier studies on cuprates \cite{DingPRL2001,FengScience2000} can also be calculated. While the gap size and the scattering rate seem almost unaffected by temperature, this later parameter exhibits a significant decrease as temperature increases towards $T_c$ \cite{ZH_LiuPRB2011}. This results suggests that the unconventional behaviour of the SC gap, with the gap filling rather than closing with temperature increasing, is associated with the lost of coherence of the quasiparticle at high-temperature, a characteristic also shared with many cuprates. 

\subsection{Scaling of the superconducting gap size with doping}

A quite trivial but yet very important remark concerning the SC gap that can be made at least for hole-doped 122-pnictides is the scaling of the SC gap size with $T_c$ as doping varies \cite{Nakayama_EPL2009, YM_Xu_UD}, as verified for all Fermi surfaces for $T_c\geq 22$ K. This result is opposite to the trend reported in some ARPES studies on cuprates and more consistent with recent works suggesting that the SC gap size in underdoped cuprates at the tip of the nodal arc scales with $T_c$ \cite{HufnerRPP2008}. Essentially, this suggests that the SC gap size is controlled by the pairing amplitude. However, such observation does not easily extend to the electron-doped side of the 122-pnictides. Indeed, a switch in the pairing amplitude ($2\Delta/k_BT_c$) from the $\alpha$ to the $\beta$ Fermi surfaces has been reported \cite{Terashima_PNAS2009}. The relative gap amplitude between the $\beta$ FS and the electronlike FSs at the M point is also different in electron-doped 122-pnictides compared to their hole-doped counter parts. Within the local antiferromagnetic exchange pairing picture, this could be explained by the fact that the gap size is not determined for each Fermi surface separately but rather depends on their momentum locations, as defined by global gap parameters determined by local antiferromagnetic exchange constants. In the relatively narrow doping range for which the scaling of the gap size with $T_c$ has been checked, the Fermi surface areas vary (and thus the SC gap size) but perhaps not enough to affect the scaling.  

\subsection{Pseudogap}

\begin{figure}[htbp]
\begin{center}
\includegraphics[width=8cm]{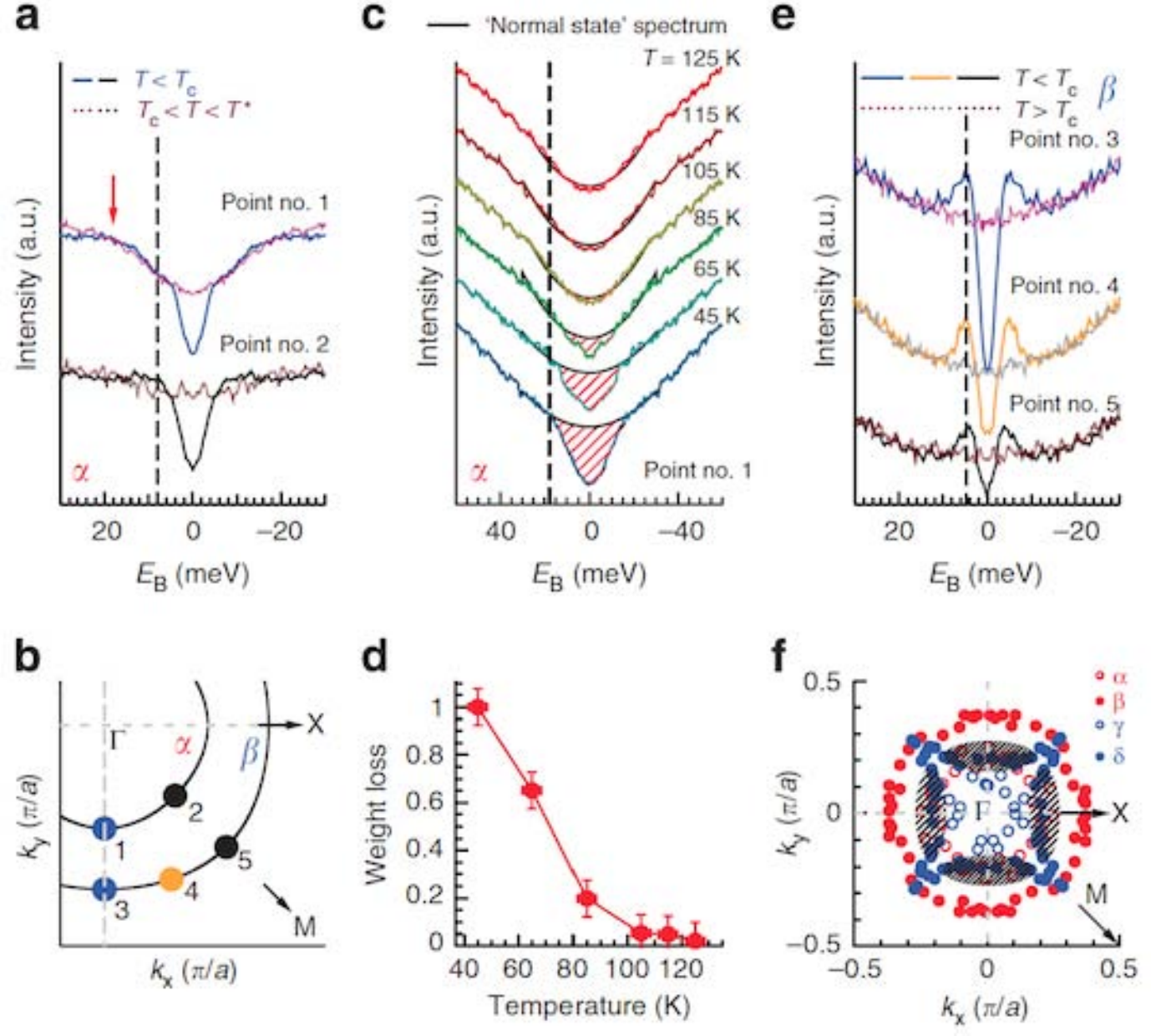}
\caption{\label{PG} (Colour online) (a) Symmetrized EDCs of Ba$_{0.75}$K$_{0.25}$Fe$_2$As$_2$ pnictide measured below and above $T_c$ at two different locations on the $\alpha$ FS (points no. 1 and no. 2 in panel (b)). The red arrow indicates that the PG is $\sim18$ meV, and the dashed vertical line shows that the SC gap on the $\alpha$ FS is $\sim8$ meV. (b) Schematic FS plot near the $\Gamma$ point indicating the measurement locations of spectra presented in (a, c and e). (c) $T$ dependence of the symmetrized EDCs of Ba$_{0.75}$K$_{0.25}$Fe$_2$As$_2$ measured at point no. 1 on the $\alpha$ FS above $T_c$. The vertical dashed line indicates the energy scale of the PG (18 meV). The shaded regions represent the spectral weight loss in the PG state. It is obtained by subtracting the symmetrized EDCs from a quadratic background. (d) T dependence of the relative weight loss (normalized by the one obtained at $T$ = 45 K) in the PG state of Ba$_{0.75}$K$_{0.25}$Fe$_2$As$_2$. The error bars represent the uncertainty in calculating the relative weight loss. (e) Similar as (a) but for the $\beta$ FS (points no. 3, no. 4 and no. 5). The dashed line shows the SC gap on the $\beta$ FS (4 meV). (f) Electron-like FS contours of Ba$_{0.75}$K$_{0.25}$Fe$_2$As$_2$ shifted to the $\Gamma$ point by the antiferromagnetic wave vector. Note that this figure is described in the 2 Fe/unit cell notation. From Xu \emph{et al.}, Nat. Commun., \textbf{2}, 392 (2011) \cite{YM_Xu_UD}, copyright \copyright\xspace (2011) by Macmillan Publishers Ltd.} 
\end{center}
\end{figure}

One of the major problems in the study of cuprates investigated by ARPES is the presence of a pseudogap \cite{DingNature1996}. Indeed, there is a suppression of density of states near the Fermi level that persists at the antinode up to a temperature $T^*$ which is much higher than $T_c$. Such behaviour is not observed near the nodes. Interestingly, a pseudogap behaviour is also observed in Ba$_{0.75}$K$_{0.25}$Fe$_2$As$_2$ \cite{YM_Xu_UD}. The corresponding feature is detected at 18 meV on the quasi-nested $\alpha$ FS. As shown in Figure \ref{PG}(a), the pseudogap feature exhibits some momentum dependence along the $\alpha$ band. While the size of the pseudogap is isotropic, the feature is more pronounced along the $\Gamma$-X direction, where the Fermi surface topology suggests a stronger antiferromagnetic scattering due to a better quasi-nesting between the $\alpha$ and $\delta$ bands, as illustrated in Figure \ref{PG}(f). Consistently, the pseudogap is absent on the unnested $\beta$ FS (see Figure \ref{PG}(e)). 

As illustrated in Figure \ref{PG}(a), the pseudogap is already present in the SC state. With temperature increasing, spectral weight lost is slowly recovered until temperature reaches $T^*=115$ K, above which no obvious change in the spectrum is detected, as shown in Figure \ref{PG}(c). The precise temperature evolution of the spectral weight loss is given in Figure \ref{PG}(d).

Combined with the evolution of the spectral lineshape with underdoping, the results of Xu \emph{et al.} on Ba$_{1-x}$K$_{x}$Fe$_2$As$_2$ indicate strong similarities with the cuprates that could origin from the importance of antiferromagnetic scattering in these two systems. In particular, the authors can identify two types of regions: i) regions where there is a smaller SC gap, no pseudogap surviving way above $T_c$ and no loss of integrity of the quasiparticle peak upon underdoping. These regions correspond to the unnested Fermi surfaces in the Fe-based superconductors and to the nodal region in the cuprates; 2) regions where the SC gap amplitude is larger and where a pseudogap surviving at high temperature is found, along with a loss of integrity of the quasiparticle peaks with underdoping. These regions correspond with regions associated with stronger antiferromagnetic scattering, namely the quasi-nested Fermi surfaces in the Fe-based superconductors and the antinodal region in the cuprates, both of which are connected by the antiferromagnetic wave vector.

\section{Discussion}
\label{section_discussion}

With its ability to \emph{resolve directly} the electronic structure, the Fermi surface and the SC gap amplitude in the momentum space, there is no doubt that ARPES is a unique and powerful tool to investigate the electronic properties of multi-band systems such as the Fe-based superconductors. Even though strictly speaking ARPES measures the electronic states in the very few layers below the surface, we presented in this review sufficient consistent results on different crystal structures and cleaved surfaces to prove empirically that ARPES is at the very least a good first order representation of the electronic states of most bulk Fe-based superconductors. For instance, the momentum dependence observed along the direction perpendicular to the cleaved surface is a strong indication that the probe states are not confined to the surface. Nevertheless, ARPES is not the only experimental technique that can be used to get some insight on the electronic behaviour of these materials and at this point we would like to make a comparison regarding the symmetry of the order parameter between ARPES and a bulk transport probe, namely thermal conductivity.  

The ARPES results on the SC order parameter are consistent, as suggested by Figure \ref{polar} for a few materials (we could also include the 111-pnictides \cite{ZH_LiuPRB2011} and the 11-chalcogenides \cite{Nakayama_PRL2010,Miao2011}): (i) the SC gap size is Fermi surface dependent; (ii) the SC gap size is in the strong coupling regime, with typical $2\Delta/k_BT_c$ ratios of 5 to 8; (iii) the SC gap size is isotropic or in the worst case nearly isotropic along each Fermi surface taking separately, except perhaps for a small warping along the $k_z$ direction, as observed in optimally-doped Ba$_{1-x}$K$_{x}$Fe$_2$As$_2$ \cite{YM_Xu_NPhys2011}. In other words, all the Fermi surfaces are fully gapped, without any nodes.

It is generally accepted by now that the SC state in the Fe-based superconductors competes with long-range antiferromagnetism. This widespread belief is reinforced by the proximity of the SC and magnetic regimes in the phase diagram. In addition, several theoretical models propose that superconductivity is mediated by short-range antiferromagnetic fluctuations. Consequently, it is very natural to expect that the application of an external magnetic field might not simply suppress superconductivity like in conventional superconductors, but also be accompanied by unwanted secondary effects such as the modulation of the antiferromagnetic fluctuations and the modification of the Fermi surface topology. For this simple reason, we prefer to first avoid to compare ARPES with experiments that necessitate the use of an external magnetic field. 

Thermal conductivity ($\kappa(T)$) is an experimental probe that does not necessitate the use of an external magnetic field. It usually contains a term linear in temperature that has an electronic origin, and a higher power law term attributed to phonons and other excitations. Near absolute zero temperature, all carriers are paired. Since Cooper pairs cannot carry entropy, the electronic term of the thermal conductivity must vanish, unless some Fermi surfaces are partly or entirely not gapped, in which case $\kappa(T=0)\neq 0$. We recall that this condition requires that the samples measured have high quality and are not phase separated. With temperature increasing, quasiparticles thermally excited above the full SC energy gap $2\Delta$ can eventually contribute to the thermal conductivity. Although thermal conductivity does not allow to directly locate nodes or non-gapped Fermi surfaces in the momentum space in the case of multiple Fermi surface sheets, it is an efficient tool to reveal the existence of such features. It is also important to add that as with ARPES, thermal conductivity in zero field does not rely on any model to claim on the presence of nodes and it is immune to localized states induced by impurities, which is a huge advantage over several other transport techniques that enhances the reliability of the conclusions.

In agreement with the result first demonstrated by ARPES experiments \cite{Ding_EPL,L_Zhao}, thermal conductivity shows a negligible $\kappa(T\rightarrow 0)/T$ term in optimally-doped Ba$_{0.6}$K$_{0.4}$Fe$_2$As$_2$, indicating that all Fermi surfaces are fully gapped \cite{XG_LuoPRB2009}. In fact, that result is still valid down to $x=0.16$ for in-plane and out-of-plane thermal conductivity in zero field \cite{Reid2011}, range that extends below the lowest $x$ value for which ARPES measurements have been performed so far. Below that range, Reid \emph{et al.} found a sudden increase in the zero field residual $\kappa(T\rightarrow 0)/T$ \cite{Reid2011}. It is still debated whether this strange behaviour is intrinsic or caused by phase separation at these low dopings. Unfortunately, considering that the gap size becomes small in this region, that the critical temperatures are low compared to typical temperatures used in ARPES and the the quasiparticle coherence weakens with underdoping \cite{YM_Xu_UD}, it is unlikely that ARPES can soon provide a definite answer on this issue. It is also not possible for ARPES to check the claim of nodes in the low-$T_c$ compound KFe$_2$As$_2$ \cite{JK_DongPRL2010} due to energy resolution and temperature limitations. However, the Fermi surface topology measured by ARPES for this system is quite different at the M point from that of the other pnictides \cite{Sato_PRL2009}, and from these results it is conceivable that this particular Fermi surface topology leads to a gap structure at the M point that includes nodes and strong anisotropy.  

Thermal conductivity measurements were also performed on the electron-doped side of the 122-pnictides. As with their hole-doped counterparts, the electron-doped 122-pnictides exhibit a negligible in-plane $\kappa(T\rightarrow 0)/T$ term \cite{L_DingNJP2009,TanatarPRL2010,ReidPRB2010}, in agreement with the nodeless SC gaps determined from ARPES measurements \cite{Terashima_PNAS2009}. The situation is different for the out-of-plane thermal conductivity, which suggests the presence of nodes \cite{ReidPRB2010}. Unfortunately, gap measurements by ARPES have never been reported so far along the $k_z$ direction for the electron-doped 122-pnictides, and a direct comparison is thus impossible. Yet, the ARPES results at a fixed $k_z$ indicate that if there are nodes in Ba$_{1}$Fe$_{2-x}$Co$_{x}$As$_2$, they must be confined to a limited $k_z$ range. 

Fe-based superconductors other than the 122-pnictides have also been investigated by thermal conductivity. For instance, Tanatar \emph{et al.} measured the in-plane and out-of-plane thermal conductivity of LiFeAs \cite{Tanatar2011}. They find no important residual $\kappa(T\rightarrow 0)/T$ term in zero field, suggesting the absence of nodes in the gap structure, in agreement with ARPES gap measurements in LiFeAs \cite{BorisenkoPRL2010} and NaFe$_{0.95}$Co$_{0.05}$As \cite{ZH_LiuPRB2011}. Similarly, Dong and coworkers found only a very small $\kappa(T\rightarrow 0)/T$ residual term in zero field measurements of FeSe$_x$ $(x\sim 1)$ by thermal conductivity \cite{JK_DongPRB2009}. The authors conclusion supports multi-gap $s$-wave superconductivity. Although the sample composition is different from that used by ARPES in Nakayama \emph{et al.} \cite{Nakayama_PRL2010} and Miao \emph{et al.} \cite{Miao2011}, the results are at least not contradictory.  

At this point we can summarize this discussion by saying that whenever ARPES and zero field thermal conductivity data have been recorded in similar conditions, the results are compatible. Now what happens when we add the magnetic field that we deliberately avoided so far to prevent any possible misinterpretation of the experimental data? Unfortunately, the interpretation of thermal conductivity in magnetic field is rarely straightforward for a multi-band system. One can compare the field dependence with that of other materials, but comparisons may be awkward if the systems have different Fermi surface topologies, which is typically the case, or if a magnetic ground state competes with superconductivity in some cases and not in the others. Alternatively, one can base his interpretation on theoretical models, with the strong assumption that these models are valid. As a rule of thumb though, it is reasonable to admit that the presence of weakly gap states will imprint a fast rising field dependence in the low temperature thermal conductivity, as these states get more easily excited.   

Although no strong SC gap anisotropy is claimed from thermal conductivity measurements in magnetic field on LiFeAs \cite{Tanatar2011} and FeSe \cite{JK_DongPRB2009}, the situation is more ambiguous for the 122-pnictides. No strong anisotropy is claimed in Ba$_{0.6}$K$_{0.4}$Fe$_2$As$_2$ for $x>0.16$, whereas nodes are suggested below that critical doping \cite{Reid2011}, which has not been investigated by ARPES. For the electron-doped side, while magnetic field dependent data on Ba$_{1}$Fe$_{1.9}$Ni$_{0.1}$As$_2$ \cite{L_DingNJP2009} and underdoped BaFe$_{2-x}$Co$_{x}$As$_2$ \cite{TanatarPRL2010} indicate fully or nearly isotropic gaps, the rapid increase in $\kappa(H)$ in the overdoped regime rather suggests a deep gap minimum somewhere on the Fermi surface \cite{TanatarPRL2010}.

In the ARPES measurements, the concepts of gap isotropy and gap anisotropy are well defined, the gap being measured directly in the momentum space around each Fermi surface taken separately. On the other hand, transport measurements do not have a real momentum resolution. It is possible, for example, to explain the thermal conductivity results in the over-doped regime of BaFe$_{2-x}$Co$_{x}$As$_2$ by claiming that the gaps around all Fermi surfaces remain isotropic while one of them, most likely associated to the vanishingly small $\beta$ Fermi surface pocket in this regime, becomes very small in a uniform fashion. Moreover, it worths to recall that unlike ARPES, transport techniques are rather sensitive to the overall gap structure around the Brillouin zone center, and depend not only on the gap size on the various Fermi surface sheets, but also on their momentum distribution with respect to the $\Gamma$ point. Hence, even though ARPES indicates isotropic SC gaps on each Fermi surface, the overall SC gap distribution with respect to the $\Gamma$ point, as measured by ARPES, is anisotropic in the Fe-based superconductors due to the Fermi surface topology. In addition, the interactions of the electronic states with an external magnetic field \emph{must} obey selection rules that reflect the point group symmetry of the crystal and that of the overall Fermi surface, which is anisotropic. 

Before concluding this section, we would like to discuss another possible source of apparent discrepancy between ARPES and some interpretations of transport measurements. As mentioned in this review, the gap $\Delta$ in ARPES measurements is usually defined as half of the full gap in the electronic dispersion below the SC transition. This gap is the same that defines the strength of the pairing interaction, and the same that characterizes the precise Bogoliubov dispersion below $T_c$. In that sense, it can arguably be called the ``true gap". However, the lifetime of the SC quasiparticles in the Fe-based superconductors, characterized by a scattering rate $\Gamma_k$, is quite short compared to that of conventional superconductors, which also means that $\Gamma_k$ is large. Consequently, there is an in-gap residual density of states whose extension may be roughly represented by the leading edge position $\delta_{LE}(k)$ of the ARPES EDCs. Assuming infinite resolution and neglecting thermal broadening, $\delta_{LE}(k)=\Delta-\Gamma_k$, which illustrates that the leading edge is necessarily an underestimation of the ``true gap" size. If ever transport measurements are sensitive to this residual density of states near $E_F$, the symmetry of the order parameter determined by these techniques would reflect the symmetry of $\delta_{LE}(k)$ rather than that of the ``true gap". Since $\Gamma_k$ is momentum-dependent and depends, for instance, on the nearly-elastic components of the antiferromagnetic scattering, we should expect a non-negligeable momentum dependence for $\delta_{LE}(k)$, especially in the presence of an applied magnetic field.

\section{Conclusion}
\label{section_conclusion}

As compared with the high-$T_c$ cuprates, the progress of our understanding of the ARPES measurements on the Fe-based superconductors during the past 4 years has been extremely fast. Interestingly, there is also much less discrepancies found in literature among work from different groups. Even though there is still no consensus on the nature of superconductivity in these materials, the contribution to the current knowledge provided by ARPES is significant and includes the precise determination of the Fermi surface topology of different compounds, the observation of electronic renormalization and band dispersion anomalies, as well as the characterization of the size and the symmetry of the SC order parameter. Thus, we can draw some very important conclusions among which are: i) the Fermi surface of the Fe-based superconductors is composed of several Fermi surface pockets; ii) the electronic dispersion of these materials is more three-dimensional than most copper oxide superconductors; iii) the band structure exhibits a non-negligible renormalization compared to LDA calculations that is more pronounced for the low-energy states near the Fermi level; iv) the SC gap of these materials is FS sheet dependent but isotropic or nearly isotropic on each FS; v) the pairing strength determined from the SC gap size indicates a pairing in the strong coupling regime; vi) antiferromagnetic scattering plays an important role in these materials. Furthermore, we provided evidence for consistency between ARPES and other types of measurement. Hopefully, the accumulation of experimental data from ARPES and other probes will soon pave the way to a global understanding of the key features leading to the exotic electronic behaviour of the Fe-based superconductors.

\section*{Acknowledgments}

P.R. and H. D. are thankful to the Chinese Academy of Sciences (grant No. 2010Y1JB6), the Ministry of Science and Technology of China (grants No. 2010CB923000, No. 2011CBA00101), and the Nature Science Foundation of China (grants No. 10974175, No. 11004232, and No. 11050110422). T.S, K.N. and T.T. are also thankful to the MEXT of Japan, JSPS, and TRIP-JST.

\section*{References}
\bibliographystyle{unsrt} 
\bibliography{biblio_en}

\end{document}